\documentclass[12pt]{article}
\usepackage{arxiv}
\usepackage[utf8]{inputenc} 
\usepackage[T1]{fontenc}  
\usepackage[tbtags]{amsmath}
\usepackage{bm}
\usepackage{graphicx}
\graphicspath{ {./Figures/} }
\usepackage{caption}
\usepackage{subcaption}
\usepackage{cool}
\usepackage[bottom]{footmisc}
\usepackage{titlesec}
\usepackage{enumitem}
\usepackage[round]{natbib}
\usepackage{hyperref}
\usepackage[capitalise]{cleveref}
\usepackage{tikz}
\usepackage{pgfplots}
\pgfplotsset{compat=1.3}
\titleformat*{\subsubsection}{\normalsize\itshape}
\usepackage{rotating}
\usepackage{array}

\title{Anisotropic dual-continuum representations for multiscale poroelastic materials: Development and numerical modelling}

\author{
  Mark Ashworth\\
  Institute of GeoEnergy Engineering\\
  Heriot-Watt Unversity, Edinburgh \\
  EH14 4AP, United Kingdom \\
  \texttt{ma174@hw.ac.uk} \\
   \And
  Florian Doster\\
  Institute of GeoEnergy Engineering\\
  Heriot-Watt Unversity, Edinburgh \\
  EH14 4AP, United Kingdom \\
  \texttt{f.doster@hw.ac.uk} \\
}

\date{}

\begin{document}
\newlength\figureheight
\newlength\figurewidth
\maketitle
\captionsetup[figure]{labelfont={bf, small},labelformat={default},labelsep=period,name={Figure}, font=small}
\setlength{\abovedisplayskip}{-5pt}

\begin{abstract}
Dual-continuum (DC) models can be tractable alternatives to explicit approaches for the numerical modelling of multiscale materials with multiphysics behaviours. This work concerns the conceptual and numerical modelling of poroelastically coupled dual-scale materials such as naturally fractured rock. Apart from a few exceptions, previous poroelastic DC models have assumed isotropy of the constituents and the dual-material. Additionally, it is common to assume that only one continuum has intrinsic stiffness properties. Finally, little has been done into validating whether the DC paradigm can capture the global poroelastic behaviours of explicit numerical representations at the DC modelling scale. We address the aforementioned knowledge gaps in two steps. First, we utilise a homogenisation approach based on Levin’s theorem to develop a previously derived anisotropic poroelastic constitutive model. Our development incorporates anisotropic intrinsic stiffness properties of both continua. This addition is in analogy to anisotropic fractured rock masses with stiff fractures. Second, we perform numerical modelling to test the dual-continuum model against fine-scale explicit equivalents. In doing, we present our hybrid numerical framework, as well as the conditions required for interpretation of the numerical results. The tests themselves progress from materials with isotropic to anisotropic mechanical and flow properties. The fine-scale simulations show anisotropy can have noticeable effects on deformation and flow behaviour. However, our numerical experiments show the DC approach can capture the global poroelastic behaviours of both isotropic and anisotropic fine-scale representations. 
\end{abstract}


\section{Introduction}
Numerical modelling of multiscale, poroelastically coupled materials can be challenging due to inherent length scale heterogeneities and multiphysics behaviours. Explicit modelling approaches allow one to account for each length scale directly within a model. This representation can therefore provide accurate and detailed descriptions. However, the number of degrees of freedom needed for direct models of multiscale, poroelastic materials can make simulation computationally prohibitive. Further, particularly within the subsurface, the data needed to populate such explicit approaches may be sparse. 

Implicit models alleviate the problems associated with explicit models, at the expense of abstraction of local scale physics. One such modelling concept is the dual-continuum (DC) model, originally attributed to \cite{Barenblatt1960}. This implicit approach has been used successfully within the context of flow modelling in a variety of subsurface engineering settings (\citealt{Gerke1993}; \citealt{Wu2002}; \citealt{Reimus2003}; \citealt{March2016}). In the DC paradigm, one continuum represents a high storage, low permeability material (e.g. matrix), whilst the other represents a low storage, high permeability material (e.g. fractures). 

This work concerns the dual-continuum modelling of multiscale, poroelastic  geomaterials. We address two knowledge gaps associated with the DC modelling paradigm within the context of poroelasticity: 

First, we develop the poroelastic DC modelling approach. We introduce the underlying modelling assumptions, whilst considering the material symmetry and mechanical properties of the constituents in the process. With respect to the latter, previous poroelastic DC models have, for the most part, assumed isotropy of the continua and bulk material (\citealt{Berryman1995}; \citealt{Khalili1996}; \citealt{Loret1999}; \citealt{Choo2015}). However, rock formations are well known to exhibit anisotropic properties (\citealt{Snow1969}; \citealt{Price1990}; \citealt{Babuska1991}). Recent work by \cite{Zhang2019} showed that anistotropic permeabilities can have measurable impacts on the flow-patterns in poroelastic dual-continuum materials. Further to anisotropy, and in the case of fractured materials, the fractures themselves can have intrinsic mechanical properties owing to local asperities and/or bridging material between fracture faces (\citealt{Olsson2001}; \citealt{Lemarchand2009}; \citealt{Jaeger2009}). Intrinsic mechanical properties of both continua have been considered for isotropic materials in the works of \cite{Elsworth1992}, \cite{Berryman2002a}, \cite{Berryman2002b} and \cite{Nguyen2010}. In this work we further explore the impact of the anisotropic elasticity, in addition to permeabilities, on dual-continuum responses.

Incorporating anisotropic and intrinsic properties can be done at the constitutive modelling stage. In the following, we add to a micromechanically derived anisotropic constitutive model by \cite{Dormieux2006}. Contrary to the model by \cite{Dormieux2006}, we incorporate linear (poro-) elastic properties for the low storage, high permeability continuum at the microscale. In this case \textit{both} continua have intrinsic stiffness properties. Following homogenisation, the resulting model, complete with expressions for the effective parameters, is an anisotropic, dual-stiffness constitutive model. Previous isotropic constitutive models reviewed in \cite{Ashworth2019b} can then be recovered under isotropy and void-space assumptions on the general anisotropic, dual-stiffness constitutive model derived herein. 

Second, with the derived poroelastic constitutive model, we proceed to numerical modelling. We investigate whether the DC representation is able to capture the global poroelastic behaviours of a fine-scale explicit model at the DC modelling scale. Whilst work has gone into testing and validating the DC concept for the flow problem (e.g. \citealt{Lewandowska2004}; \citealt{Egya2019}), little has been done to asses validity of the poroelastically coupled DC approach. Further, we discuss several considerations to ensure meaningful interpretations between the two modelling approaches. 

To summarise, our aims are twofold. First, in Section 2, we use a homogensation approach and develop a previously introduced anisotropic dual-continuum constitutive model. For this development we allow both continua to have (anisotropic) intrinsic mechanical properties. Second, we perform numerical modelling of the poroelastic dual-continuum concept, investigating its validity against fine-scale explicit representations. To do so we introduce the hybrid numerical framework used to perform the numerical study in Section 3. We present the numerical tests, modelling considerations and test results in Section 4. For our study we consider numerical test cases as conceptualisations of naturally fractured rock samples that satisfy certain representative elementary volume (REV) requirements. Our results show that the DC model is capable of capturing the global poroelastic behaviours of isotropic and anisotropic fine-scale equivalents. Finally, we offer conclusions and recommendations for future work in Section 5.

\section{Homogenisation of the dual problem}
In the following we develop the anisotropic, dual-stiffness constitutive model for a poroelastic DC material. To do so we expand the homogenisation approach originally proposed by \cite{Dormieux2006} by including intrinsic mechanical properties for the low storage, high permeability continuum. 

Whilst this work strictly assumes linear poroelasticity, the inclusion of stiffness properties are necessary for extensions to non-linear modelling of materials (\citealt{borja2016}). For example, it is well known that mechanically weak materials, such as fractures, show non-linearly elastic, or inelastic, hardening behaviours even at small deformations (\citealt{Bemer2001}; \citealt{Deude2002}; \citealt{Lemarchand2009}; \citealt{Bidgoli2013}). We acknowledge the simplifying assumptions used in the current work, with a view to incorporating more realistic deformation behaviours on the basis of the modelling concepts developed herein.

To keep notation brief, we refer to the low permeability storage continuum as the \textit{matrix} continuum, and the low storage, high permeability transport continuum as the \textit{fracture} continuum. However, this work is sufficiently general such that other multiscale materials can be considered e.g. soil aggregates (under the assumption of infinitesimal deformations) (\citealt{Choo2015}; \citealt{Choo2016}).  

\subsection{Volume averaging}
\begin{figure}[h!]
\centering
\includegraphics[scale =0.90]{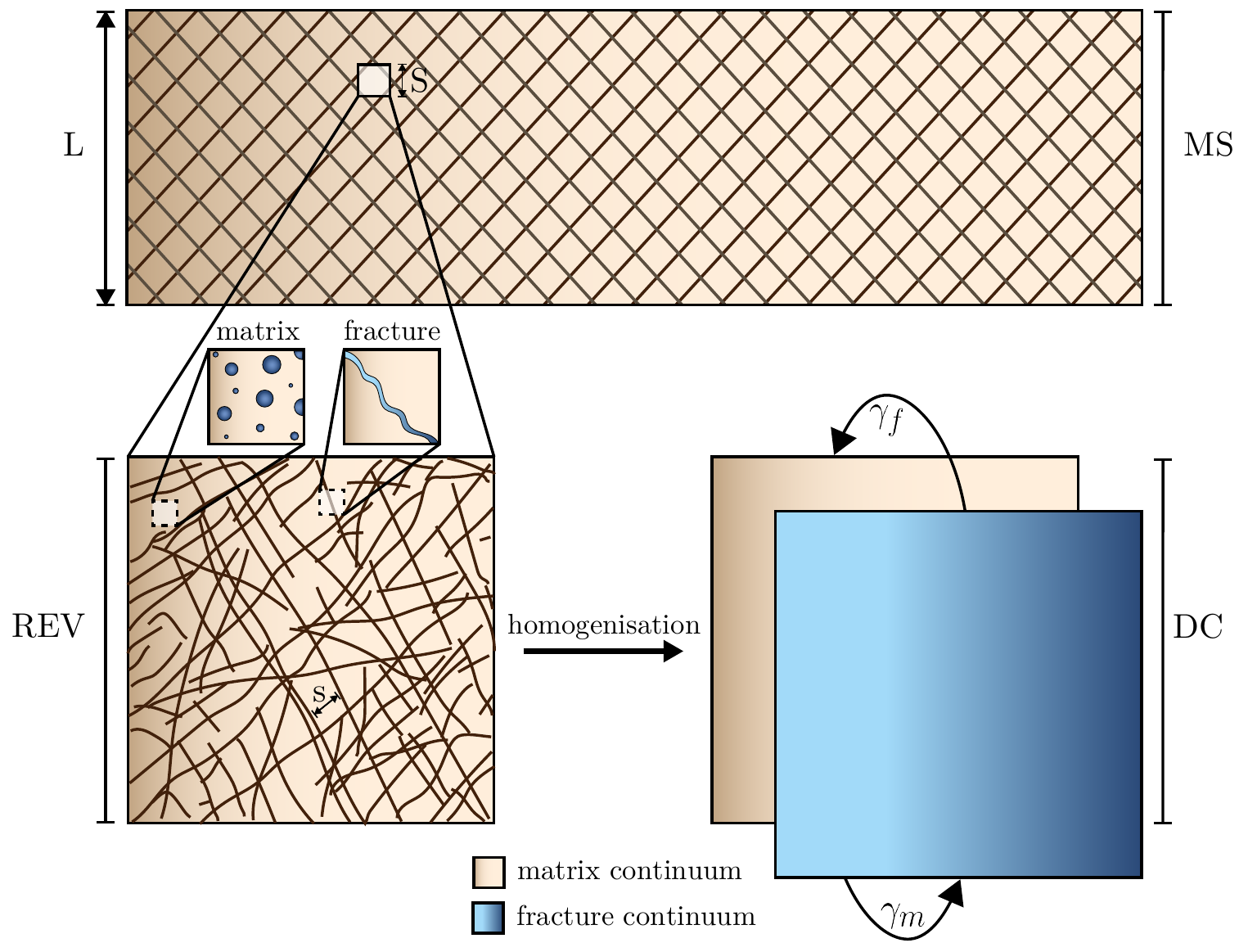}
\caption{A geometrical interpretation of an REV over a microscopic scale from a large macroscopic structure (MS). The REV is used to define the macroscopic dual-continuum (DC) model in which matrix ($m$) and fracture continua ($f$) are superposed in space and time. Inter-continuum mass exchange is described by the transfer term $\gamma_\alpha$ $[\alpha = m,f]$. Notations $s$, $S$ and $L$ denote characteristic lengths of local heterogeneities, the REV and the macroscopic structure respectively.} \label{fig:1}
\end{figure}

We define the averaging operation, and assumptions therein, required in the homogenisation approach. A dual-continuum representation can be justified if an REV can be taken from a large macroscopic structure. Identification of an REV requires the satisfaction of the scale separation principle summarised as (\citealt{Bear2012}),
\begin{equation}
    s \ll S \ll L. \label{eqn:1}
\end{equation} 
where $s$, $S$ and $L$ denote the characteristic lengths at the local heterogeneity, REV and macroscopic body scales respectively. \cref{eqn:1} should honour length scale requirements for the physical system, both geometrically (\cref{fig:1}), and with respect to the wavelengths of the physical process (\citealt{Auriault2002}; \citealt{Geers2010}). Accordingly, the REV represents the scale at which relationships between averaged quantities are defined (\cref{fig:1}). 

Defining an REV over fractured media is a subject of much debate due to the challenge of establishing criteria for scale separation (\citealt{Long1982}; \citealt{Neuman1988}; \citealt{Min2003}; \citealt{Berre2019}). In the following, however, we suppose a material for which an REV can be defined, such as densely fractured rock masses (\citealt{Berkowitz2002}).

To proceed we assume statistical homogeneity of the underlying material and thus make use of the volume average over the REV (\citealt{Nemat1993}),
\begin{equation}
    \overline{\bm{z}} = \frac{1}{|\Omega|}\int_{\Omega}\bm{z}(\bm{x})\text{ } dV, \label{eqn:2}
\end{equation}
where $\bm{z}$ is an arbitrary tensor field, $\bm{x}$ is a position vector locating an REV within the macroscopic composite body, and $|\Omega|$ is the volume of an arbitrary REV within the body. 

\subsection{Homogenisation}
\subsubsection{Preliminaries}
To develop the homogenisation problem introduced by \cite{Dormieux2006} we consider the domain, $\Omega$, over an REV in which there exists a porous matrix continuum, $\Omega_m \subset \Omega$, and porous fracture continuum $\Omega_f \subset \Omega$. We assume linear poroelasticity for each continuum (\citealt{Biot1941}, \citealt{Coussy2004}), and that the continua are saturated by the same slightly compressible fluid. Further, we assume isothermal evolutions and zero initial stress and pressure conditions. 

An important assumption is that we consider microscopic fluctuations in pressures are negligible with respect to the macroscopic (average) continuum pressures (\citealt{Dormieux2006}). As a result, fluids are assumed to be in steady state, but at different equilibrium pressures, within the respective continua in the REV. Accordingly, we model solid-fluid interactions at the microscale using uniform continuum pressures (\citealt{Van2016}). 

With the given assumptions, the local constitutive model for a continuum, $\alpha$, is then
\begin{align}
    \bm{\sigma}_\alpha = \mathbb{C}_\alpha:\bm{\epsilon}_\alpha - \bm{b}_\alpha P_\alpha &\quad \text{in } \Omega_\alpha, \label{eqn:3} \\
    \text{d}\varphi_\alpha = \bm{b}_\alpha:\bm{\epsilon}_\alpha + \frac{1}{n_\alpha}P_\alpha &\quad \text{in } \Omega_\alpha. \label{eqn:4}
\end{align}
where $\mathbb{C}_\alpha$ $[\alpha = m,f]$ is the intrinsic fourth-order stiffness tensor for continuum $\alpha$, and the second-order tensors, $\bm{\sigma}_\alpha, \text{ } \bm{\epsilon}_\alpha, \text{ } \bm{b}_\alpha$, are the microscopic Cauchy stress and linearised strain tensors, and intrinsic Biot coefficient for continuum $\alpha$ respectively. Parameter $n_\alpha^{-1}$ is the inverse of the Biot modulus, $P_\alpha$ is the macroscopic fluid pressure, and $\text{d}\varphi_\alpha=\varphi_\alpha-\varphi_\alpha^0$ is the evolution of the local Lagrangian porosity from the reference state (denoted by superscript $0$), all written in terms of continuum $\alpha$. The local Lagrangian porosity is the ratio of the continuum pore volume $|\Omega_\alpha^p|$, to the bulk volume of the undeformed continuum configuration, $|\Omega_\alpha^0|$. As is customary, we take stress and strain as positive in the tensile direction. 

It is useful to re-write \crefrange{eqn:3}{eqn:4} in a unified way as follows (\citealt{Dormieux2006}),
\begin{equation}
    \bm{\sigma}(\bm{x}) = \mathbb{C}(\bm{x}):\bm{\epsilon}(\bm{x}) + \bm{\sigma}^p(\bm{x}) \quad \forall \bm{x} \in\Omega, \label{eqn:5}
\end{equation}
where $\mathbb{C}(\bm{x})$, and the prestress tensor distribution related to the fluid pressure (\citealt{Chateau2002}), $\bm{\sigma}^p(\bm{x})$, are given by
\begin{equation}
    \mathbb{C}(\bm{x}) = 
        \begin{cases}
            \mathbb{C}_m &\text{in } \Omega_m \\
            \mathbb{C}_f &\text{in } \Omega_f 
        \end{cases}, \label{eqn:6}
\end{equation}
\begin{equation}
    \bm{\sigma}^{p}(\bm{x}) = 
        \begin{cases}
            -\bm{b}_mP_m &\text{in } \Omega_m \\
            -\bm{b}_fP_f &\text{in } \Omega_f 
        \end{cases}, \label{eqn:7}
\end{equation}
respectively.

The essence of the homogenisation approach is to define a boundary value problem on the REV, the solution to which allows for the determination of macroscopic constitutive properties. Accordingly the conservation of momentum boundary value problem is defined as
\begin{align}
    \nabla\cdot\bm{\sigma} = 0 &\quad \text{in } \Omega, \label{eqn:8} \\
    \bm{\sigma} = \mathbb{C}(\bm{x}):\bm{\epsilon} + \bm{\sigma}^p(\bm{x}) &\quad \text{in } \Omega, \label{eqn:9} \\
    \hat{\bm{u}} = \textbf{E}\cdot\bm{x} &\quad \text{on } \partial\Omega, \label{eqn:10}
\end{align}
where $\partial\Omega$ is the boundary of $\Omega$, $\bm{u}$ is the microscopic displacement vector and $\textbf{E}$ is the macroscopic (or surface prescribed) strain tensor. Quantities denoted by $\text{ }\hat{}\text{ }$ are boundary assigned values i.e. $\bm{u}=\hat{\bm{u}}$ on $\partial\Omega$. 

Using the averaging operation, \cref{eqn:2}, it can be shown (e.g. \citealt{Hashin1972}) for uniform displacement boundary conditions, \cref{eqn:10}, that
\begin{equation}
    \textbf{E} = \overline{\bm{\epsilon}}, \label{eqn:11}
\end{equation}
and
\begin{equation}
    \overline{\bm{\epsilon}} = v_m\overline{\bm{\epsilon}}_m + v_f\overline{\bm{\epsilon}}_f, \label{eqn:12}
\end{equation}
where $\overline{\bm{z}}_\alpha=|\Omega_\alpha|^{-1}\int_{\Omega_\alpha}\bm{z}_\alpha(\bm{x})\text{ }dV$, and $v_\alpha$ is the volume fraction of continuum $\alpha$, defined as the ratio of the continuum volume, $|\Omega^{0}_\alpha|$, to the bulk volume, $|\Omega^0|$, taken at reference conditions.

To give the link between microscopic fields, in this case strain, and macroscopic counterparts we consider a mapping between $\epsilon(\bm{x})$ and $\textbf{E}$. Owing to the linearity of \cref{eqn:8} we can define a linear mapping so that
\begin{equation}
    \bm{\epsilon}(\bm{x}) = \mathbb{A}(\bm{x}):\textbf{E}. \label{eqn:13}
\end{equation}
where $\mathbb{A}(\bm{x})$ is the fourth-order mapping tensor (\citealt{Hill1963}). 

Finally, combining \crefrange{eqn:11}{eqn:13} it can be shown  
\begin{equation}
    \overline{\bm{\epsilon}} = v_m\overline{\mathbb{A}}_m:\textbf{E} + v_f\overline{\mathbb{A}}_f:\textbf{E}, \label{eqn:14}
\end{equation}
from which we can see
\begin{equation}
     \mathbb{I} = \overline{\mathbb{A}} = v_m\overline{\mathbb{A}}_m + v_f\overline{\mathbb{A}}_f \label{eqn:15}
\end{equation}
where $\mathbb{I}$ is the fourth-order identity tensor.

\subsubsection{Recovery of the constitutive system}
From the superposition property in linear systems, \crefrange{eqn:8}{eqn:10} can be decomposed into two subproblems. Subproblem I can be interpreted as a drained poroelastic problem:
\begin{align}
    \nabla\cdot\bm{\sigma}^\text{I} = 0 &\quad \text{in } \Omega, \label{eqn:16} \\
    \bm{\sigma}^\text{I} = \mathbb{C}(\bm{x}):\bm{\epsilon}^\text{I} &\quad \text{in } \Omega, \label{eqn:17} \\
    \hat{\bm{u}}^\text{I} = \textbf{E}\cdot\bm{x} &\quad \text{on } \partial\Omega, \label{eqn:18} \\
    \text{with} \qquad \bm{\Sigma}^\text{I} = \overline{\bm{\sigma}}^\text{I} = \mathbb{C}^*:\textbf{E} &\quad  \label{eqn:19} 
\end{align}
where the macroscopic stress tensor, $\bm{\Sigma}=\overline{\bm{\sigma}}$ (\citealt{Hashin1972}), and $\mathbb{C}^*=\overline{\mathbb{C}:\mathbb{A}}:\textbf{E}$ is the upscaled stiffness tensor for the dual-material. Subproblem II defines a constrained material, $\textbf{E}=0$, subject to loading via the prestress field, $\bm{\sigma}^p$:
\begin{align}
    \nabla\cdot\bm{\sigma}^\text{II} = 0 &\quad \text{in } \Omega, \label{eqn:20} \\
    \bm{\sigma}^\text{II} = \mathbb{C}(\bm{x}):\bm{\epsilon}^\text{II} + \bm{\sigma}^p(\bm{x}) &\quad \text{in } \Omega, \label{eqn:21} \\
    \hat{\bm{u}}^\text{II} = 0 &\quad \text{on } \partial\Omega, \label{eqn:22} \\
    \text{with} \qquad \bm{\Sigma}^\text{II} = \overline{\bm{\sigma}}^\text{II} = \overline{\mathbb{C}(\bm{x}):\bm{\epsilon}^\text{I}} + \overline{\bm{\sigma}^p(\bm{x})} &\quad \label{eqn:23} 
\end{align}
From \cite{Dormieux2006} one can show
\begin{equation}
    \bm{\Sigma}^\text{II} = \bm{\Sigma}^p, \label{eqn:24}
\end{equation}
where $\bm{\Sigma}^p=\overline{\bm{\sigma}^p:\mathbb{A}}$. \cref{eqn:24} is a part of a classical result in micromechanics referred to as Levin's theorem (\citealt{Levin1967}). That is, the macroscopic constitutive equation follows the form of the linear local constitutive relation, \cref{eqn:3},
\begin{equation}
    \bm{\Sigma} = \bm{\Sigma}^\text{I} + \bm{\Sigma}^\text{II} = \mathbb{C}^*:\textbf{E} + \bm{\Sigma}^p, \label{eqn:25}
\end{equation}
where we make use of the linearity of the problem to superpose subproblems I and II. Owing to the definition of $\mathbb{C}(\bm{x})$, and from \crefrange{eqn:14}{eqn:15}, the homogenised stiffness tensor of the composite dual-material is defined as
\begin{equation}
    \mathbb{C}^* = v_m\overline{\mathbb{A}}_m:\mathbb{C}_m + (\mathbb{I}-v_m\overline{\mathbb{A}}_m):\mathbb{C}_f. \label{eqn:26}
\end{equation}
Similarly, the homogenised prestress tensor is given as
\begin{equation}
    \bm{\Sigma}^p = -v_m\overline{\mathbb{A}}_m:\bm{b}_mP_m - (\mathbb{I} - v_m\overline{\mathbb{A}}_m):\bm{b}_fP_f. \label{eqn:27}
\end{equation}
Intuitively, \cref{eqn:27} can be interpreted as a weighted sum of the continuum pressures. In the work of \cite{Borja2009}, the authors derive a pore fraction weighting formulation that is thermodynamically consistent. Such an approach was also proposed in \cite{Coussy2004}. Given the thermodynamic consistency, it would be interesting to see how one could recover a pore fraction weighted formulation within the general framework of microporomechanics. 

To proceed, using \cref{eqn:27}, and with the result from \cref{eqn:26}, we can identify the first of the macroscopic constitutive parameters, that is the effective Biot coefficients,
\begin{gather}
    \bm{B}_m = \bm{b}_m:\left[\left(\mathbb{C}^*-\mathbb{C}_f\right):\left({\mathbb{C}_m-\mathbb{C}_f}\right)^{-1}\right],  \label{eqn:28} \\
    \bm{B}_f = \bm{b}_f:\mathbb{I} - \bm{b}_f:\left[\left(\mathbb{C}^*-\mathbb{C}_f\right):\left({\mathbb{C}_m-\mathbb{C}_f}\right)^{-1}\right].  \label{eqn:29}
\end{gather}
From the energy approach to poromechanics (\citealt{Coussy2004}), the dual-continuum model requires state equations for the evolutions of macroscopic Lagragian porosity (\citealt{Ashworth2019b}). Accordingly, for subproblem I 
\begin{align}
    \text{d}\phi^\text{I}_\alpha = v_\alpha\overline{\text{d}\varphi}_\alpha &= v_\alpha\overline{\mathbb{A}}_\alpha:\bm{b}_\alpha:\textbf{E} \nonumber \\
    &= \bm{B}_\alpha:\textbf{E}. \label{eqn:30}
\end{align}
where we have made use of \cref{eqn:4} in defining \cref{eqn:30}. 

Given subproblem II we have 
\begin{align}
    \text{d}\phi^\text{II}_m &= v_m\bm{b}_m:\overline{\bm{\epsilon}}^\text{II}_m + \frac{v_m}{n_m}P_m, \label{eqn:31} \\
    \text{d}\phi^\text{II}_f &= -v_m\bm{b}_f:\overline{\bm{\epsilon}}^\text{II}_m + \frac{v_f}{n_f}P_f, \label{eqn:32}
\end{align}
where we have used \cref{eqn:11} together with the fact $\overline{\bm{\epsilon}}^\text{II}=0$ to eliminate $v_f\overline{\bm{\epsilon}}_f$. To advance we must substitute for $v_m\overline{\bm{\epsilon}}^\text{II}_m$. Following \citealt{Dormieux2006}, $v_m\overline{\bm{\epsilon}}^\text{II}_m$ can be expressed as
\begin{equation}
    v_m\overline{\bm{\epsilon}}^\text{II}_m = \left(\mathbb{C}_m-\mathbb{C}_f\right)^{-1}:[(v_m\bm{b}_m - \bm{B}_m)P_m + (v_f\bm{b}_f - \bm{B}_f)P_f]. \label{eqn:33}
\end{equation}
With \cref{eqn:33} in \crefrange{eqn:31}{eqn:32} we recover 
\begin{align}
    \text{d}\phi^\text{II}_m &= \frac{1}{N_m}P_m + \frac{1}{Q_m}P_f, \label{eqn:34} \\
    \text{d}\phi^\text{II}_f &= \frac{1}{Q_f}P_m + \frac{1}{N_f}P_f, \label{eqn:35}
\end{align}
where the effective constitutive parameters $N^{-1}_\alpha$ and $Q^{-1}_\alpha$ are defined as
\begin{align}
    \frac{1}{N_m} &= \bm{b}_m:\left[(v_m\bm{b}_m-\bm{B}_m):\left(\mathbb{C}_m-\mathbb{C}_f\right)^{-1}\right]+\frac{v_m}{n_m}, \label{eqn:36} \\
    \frac{1}{Q_m} &= \bm{b}_m:\left[(v_f\bm{b}_f-\bm{B}_f):\left(\mathbb{C}_m-\mathbb{C}_f\right)^{-1}\right], \label{eqn:37} \\
    \frac{1}{N_f} &= \bm{b}_f:\left[(\bm{B}_f-v_f\bm{b}_f):\left(\mathbb{C}_m-\mathbb{C}_f\right)^{-1}\right]+\frac{v_f}{n_f}, \label{eqn:38} \\
    \frac{1}{Q_f} &= \bm{b}_f:\left[(\bm{B}_m-v_m\bm{b}_m):\left(\mathbb{C}_m-\mathbb{C}_f\right)^{-1}\right]. \label{eqn:39}
\end{align}
Provided the storage continuum is isotropic, $Q^{-1}_m=Q^{-1}_f$ since $\bm{b}_m=b_m\bm{1}$ where $\bm{1}$ is the second-order identity tensor (\citealt{Dormieux2006}). 

Finally, through superposition of subproblems I and II for the macroscopic variables $\bm{\Sigma}$ and $\text{d}\phi_\alpha$, we recover the anisotropic, dual-stiffness constitutive model for the dual-scale, poroelastic material as
\begin{align}
    \bm{\Sigma} &= \mathbb{C}^*:\textbf{E} - \bm{B}_mP_m - \bm{B}_fP_f, \label{eqn:40} \\
    \text{d}\phi_m &= \bm{B}_m:\textbf{E} + \frac{1}{N_m}P_m + \frac{1}{Q_m}P_f, \label{eqn:41} \\
     \text{d}\phi_f &= \bm{B}_f:\textbf{E} + \frac{1}{Q_f}P_m + \frac{1}{N_f}P_f, \label{eqn:42} 
\end{align}
where expressions for the effective constitutive parameters $\mathbb{C}^*$, $\bm{B}_m$, $\bm{B}_f$, $N_m^{-1}$, $Q_m^{-1}$, $N_f^{-1}$, and $Q_f^{-1}$ are given by \cref{eqn:26}, \crefrange{eqn:28}{eqn:29}, and \crefrange{eqn:36}{eqn:39} respectively. 

\subsubsection{Model equivalencies}
Under certain conditions, the parameter models just referenced reduce to other mechanical property based parameter models proposed in literature. For example, in the case of soils, the high permeability (transport) continuum is all void space (\citealt{Koliji2008}). As a result $\mathbb{C}_f=0$, and we recover the original anisotropic parameter models proposed by \cite{Dormieux2006}. For an isotropic material, the constitutive system can be written in terms of scalar invariants of the tensorial quantities. Accordingly, with a change from the mixed-compliance constitutive formulation to a pure-stiffness formulation, we recover the dual-stiffness models introduced by \cite{Berryman2002a} (\citealt{Ashworth2019b}). \cite{Ashworth2019b}, using parameter models by \cite{Berryman2002a}, show some possible situations in which the inclusion of intrinsic fracture stiffness properties may be important. Further, in \cite{Kim2012}, the authors show use of \cite{Berryman2002a} type coefficient models lead to well-posed mathematical problems, an important consideration for numerical modelling. Finally, combining the void space transport and isotropic dual-material assumptions, allows us to recover parameter models originally proposed by \cite{Berryman1995} and \cite{Khalili1996}.

Under long-term drainage, $P_m=P_f$, dual-continuum models should reduce to single-continuum equivalents (\citealt{Berryman1995}). As a result, we recover the following compatibility relations:
\begin{equation}
    \bm{B} = \bm{1}-\mathbb{C}^*:\mathbb{C}^{-1}_s:\bm{1} = \bm{B}_m + \bm{B}_f, \label{eqn:43}
\end{equation}
\begin{equation}
    \frac{1}{N} = \left(\bm{B}-\phi\bm{1}\right):\mathbb{C}^{-1}_s:\bm{1}= \frac{1}{N}_m + \frac{1}{N}_f + \frac{1}{Q}_m + \frac{1}{Q}_f. \label{eqn:44}
\end{equation}
where $\mathbb{C}_s$ is the solid-grain stiffness tensor and $\phi=\phi_m+\phi_f$. \crefrange{eqn:43}{eqn:44} hold provided $\mathbb{C}_s$ is the same for both the matrix and fracture continua. Accordingly, applying the long-term drainage condition, and contracting \crefrange{eqn:40}{eqn:42} we recover the single-porosity constitutive model originally proposed in \cite{Biot1941}, albeit for anisotropic materials. Alternatively, we could recover the single-porosity model by setting $v_f=0$, and thus $\mathbb{C}_f=0$ with $\mathbb{C}^*=\mathbb{C}_m$. 

\section{Numerical framework}
Here we introduce the computational framework used for modelling the coupled DC problem. We start by introducing the strong form of the DC poroelastic problem and then progressing to its fully discrete counterpart.

\subsection{Strong form}
In addition to \crefrange{eqn:40}{eqn:42}, we require constitutive relations for intra- and inter-continuum mass flux terms, $\bm{w}_\alpha$ and $\gamma_\alpha$ respectively. Intra-continuum mass flux is given according to Darcy's law,
\begin{equation}
\bm{w}_\alpha = \rho_{l}\bm{q}_\alpha = -\rho_{l}\frac{\bm{k}^*_\alpha}{\mu_l}\cdot(\nabla P_\alpha-\rho_{l}\bm{g}), \label{eqn:45}
\end{equation}
where $\bm{q}_\alpha$ is the volumetric flux vector associated with continuum $\alpha$, $\rho_l$ and $\mu_l$ are the intrinsic fluid density and fluid viscosity respectively, $\bm{g}$ is the gravity vector and $\bm{k}^*_\alpha$ is the macroscopic continuum permeability tensor. The inter-continuum mass flux, $\gamma_\alpha$, is given according to a first-order transfer term originally proposed by \cite{Warren1963},
\begin{equation}
    \gamma_\alpha = \rho_l\frac{\varkappa k'}{\mu_l}(P_\beta - P_\alpha), \label{eqn:46}
\end{equation}
where $k'$ denotes the interface permeability, taken here as the intrinsic matrix permeability (\citealt{Barenblatt1960}; \citealt{Choo2015}), and $\varkappa$ is a parameter referred to as the shape factor (\citealt{Warren1963}). In this work we use an analytically derived $\varkappa$ for an isotropic matrix given according to \cite{Lim1995},
\begin{equation}
    \varkappa = \frac{N\pi^2}{s^2}, \label{eqn:47}
\end{equation}
where $N$ is a dimension parameter related to the number of fracture sets, and $s$ is the characteristic spacing length of the fracture continuum (\cref{fig:1}). 

Finally, we give the compatibility between the macroscopic linearised strain tensor and macroscopic displacement, $\bm{U}$, as
\begin{equation}
    \textbf{E} = \underline{\nabla}\bm{U} = \frac{1}{2}(\nabla\bm{U} + \nabla^{\top}\bm{U}), \label{eqn:48}
\end{equation}
where we introduce notation $\underline{\nabla}$ to denote the symmetric gradient operator on $\bm{U}$. 

The conservation equations considered for the DC poroelastic problem are the momentum equation
\begin{equation}
    \nabla\cdot\bm{\Sigma} + \rho\bm{g} = \tilde{\gamma}\label{eqn:49}
\end{equation}
and the continuity equations for each continuum
\begin{align}
    \pderiv{m_{l,m}}{t} + \nabla\cdot\bm{w}_m&=\gamma_m, \label{eqn:50} \\ 
    \pderiv{m_{l,f}}{t} + \nabla\cdot\bm{w}_f&=\gamma_f. \label{eqn:51}
\end{align}
Notations $\rho$ and $\tilde{\gamma}$ in \cref{eqn:49} are the bulk density of the dual-material, and a momentum source arising from the inter-continuum mass transfer respectively. For the remainder we assume $\tilde{\gamma}\approx0$, with respect to the other force density terms in \cref{eqn:49}. Notation $m_{l,\alpha}$ in \crefrange{eqn:50}{eqn:51} is the fluid mass content associated with continuum $\alpha$. The fluid mass content is given by $m_{l,\alpha}=\rho_l\phi_\alpha$. 

We consider the conservation equations over a domain, $\Omega^\mathcal{D} \subset \mathbb{R}^2$, bounded by $\partial\Omega^\mathcal{D}$. The domain boundary is separated into disjoint boundary segments corresponding to Dirichlet and Neumann boundary conditions for the mechanical and flow problems. For the mechanical problem this implies displacement ($\Gamma^{\text{U}}$) and traction ($\Gamma^{\text{T}}$) boundary conditions. To ensure well-posedness $\Gamma^{\text{U}} \cup \Gamma^{\text{T}} = \partial\Omega^\mathcal{D}$ and $\Gamma^{\text{U}} \cap \Gamma^{\text{T}} = \varnothing$. For the flow problem the boundary conditions for a given continuum are pressure ($\Gamma^{\text{P}}_\alpha$) and flux ($\Gamma^{\text{Q}}_\alpha$). Again, for a well-posed problem we have $\Gamma^{\text{P}}_\alpha \cup \Gamma^{\text{Q}}_\alpha = \partial\Omega^\mathcal{D}$ and $\Gamma^{\text{P}}_\alpha \cap \Gamma^{\text{Q}}_\alpha = \varnothing$. 

The strong form is finally defined as: Find $\bm{U}$, $P_m$ and $P_f$ that satisfy \crefrange{eqn:49}{eqn:51} subject to boundary conditions:
\begin{align}
    \bm{U} = \hat{\bm{U}}  &\quad \text{on } \Gamma^{\text{U}}, \label{eqn:52}\\
    \bm{\Sigma}\cdot\bm{n} = \hat{\bm{T}}  &\quad \text{on } \Gamma^{\text{T}}, \label{eqn:53} \\
    P_m = \hat{P}_m  &\quad \text{on } \Gamma^{\text{P}}_m, \label{eqn:54} \\
    \bm{q}_m\cdot\bm{n} = \hat{q}_m  &\quad \text{on } \Gamma^{\text{Q}}_m, \label{eqn:55} \\
    P_f = \hat{P}_f  &\quad \text{on } \Gamma^{\text{P}}_f, \label{eqn:56} \\
    \bm{q}_f\cdot\bm{n} = \hat{q}_f  &\quad \text{on } \Gamma^{\text{Q}}_f, \label{eqn:57}
\end{align}
with initial conditions
\begin{equation}
    \bm{U} = \bm{U}^0, \quad P_m = P^0_m, \quad P_f = P^0_f, \label{eqn:58}
\end{equation}
for all $(\bm{X},t)\in(\Omega^{\mathcal{D}}\times t=0)$. Notation $\bm{X}$ is the macroscopic position vector.

The single porosity linear poroelastic model can be recovered from \crefrange{eqn:49}{eqn:58} under the assumption $P=P_m=P_f$ and combining \cref{eqn:50} and \cref{eqn:51}. With the contraction to a single continuum system, DC constitutive parameters reduce to single porosity equivalents, \crefrange{eqn:43}{eqn:44}.

\subsection{Weak form}
The weak formulation of the strong form introduced previously requires the definition of the appropriate function spaces. Accordingly, solution spaces for continuum pressure and the displacements are $\mathcal{S}_{P_\alpha} = L^2(\Omega^{\mathcal{D}})$ and $\mathcal{S}_{U} = \{ \bm{U}\in H^1(\Omega^{\mathcal{D}})^d:\bm{U} = \hat{\bm{U}} \text{ on } \Gamma^{\text{U}}\}$ respectively, where $L^2$ and $H^1$ are the typical square integrable and first-order Sobolev function spaces. Weighting function spaces are then defined as $\mathcal{W}_{P_\alpha} = L^2(\Omega^{\mathcal{D}})$ and $\mathcal{W}_{U} = \{ \bm{\eta}\in H^1(\Omega^{\mathcal{D}})^d:\bm{\eta} = \bm{0} \text{ on } \Gamma^{\text{U}}\}$.

To progress we substitute the constitutive equations, \crefrange{eqn:40}{eqn:42} and \crefrange{eqn:45}{eqn:46}, and macroscopic strain compatibility relation, \cref{eqn:48}, into \crefrange{eqn:49}{eqn:51}. We adopt the material and fluid assumptions introduced in Section 2, whilst also neglecting gravitational effects. Assuming isotropic matrix material results in $Q^{-1}_m=Q^{-1}_f=Q^{-1}$ and $\bm{B}_m=B_m\bm{1}$. Further, we restrict our anisotropic experiments to orthotropic materials. Finally, comparing trial functions against weight functions, the weak form is defined as: Find $(\bm{U}, P_m, P_f) \in (\mathcal{S}_{U} \times \mathcal{S}_{P_m} \times \mathcal{S}_{P_f})$ such that for all $(\bm{\eta}, \omega_m, \omega_f) \in (\mathcal{W}_{U} \times \mathcal{W}_{P_m} \times \mathcal{W}_{P_f})$
\begin{align}
    g(\bm{\eta}, &\bm{U}) - \int_{\Omega^{\mathcal{D}}} (\nabla\bm{\eta})\cdot\bm{B}_mP_m \text{ d}V - \int_{\Omega^{\mathcal{D}}} (\nabla\bm{\eta})\cdot\bm{B}_fP_f \text{ d}V \nonumber \\ 
    &= \int_{\Gamma^{\text{T}}} \bm{\eta}\cdot\hat{\bm{T}} \text{ d}S, \label{eqn:59} \\
    \int_{\Omega^{\mathcal{D}}} &\pderiv{}{t}\omega_m\left(\bm{B}_m:\underline{\nabla}\bm{U} + \frac{1}{M_m}P_m + \frac{1}{Q}P_f\right) \text{ d}V \nonumber \\  &-\int_{\Omega^{\mathcal{D}}}\omega_m\nabla\cdot\left(\frac{\bm{k}^*_m}{\mu_l}\cdot\nabla P_m\right) \text{ d}V = \int_{\Omega^{\mathcal{D}}}\omega_m\frac{\varkappa k'}{\mu_l}(P_f - P_m) \text{ d}V, \label{eqn:60} \\
    \int_{\Omega^{\mathcal{D}}} &\pderiv{}{t}\omega_f\left(\bm{B}_f:\underline{\nabla}\bm{U} + \frac{1}{Q}P_m + \frac{1}{M_f}P_f\right) \text{ d}V \nonumber \\
    &- \int_{\Omega^{\mathcal{D}}}\omega_f\nabla\cdot\left(\frac{\bm{k}^*_f}{\mu_l}\cdot\nabla P_f\right) \text{ d}V = \int_{\Omega^{\mathcal{D}}}\omega_f\frac{\varkappa k'}{\mu_l}(P_m - P_f) \text{ d}V,  \label{eqn:61} 
\end{align}
The bilinear form $g(\cdot,\cdot)$ in \cref{eqn:59} is given by
\begin{equation}
    g(\bm{\eta}, \bm{U}) = \int_{\Omega^{\mathcal{D}}} \underline{\nabla}\bm{\eta}:\bm{\Sigma}'(\bm{U}) \text{ d}V \label{eqn:62} 
\end{equation}
where $\bm{\Sigma}'(\bm{U})=\mathbb{C}^*:\underline{\nabla}\bm{U}$ is the effective stress tensor. The term $M^{-1}_\alpha$ in \crefrange{eqn:60}{eqn:61} is given as 
\begin{equation}
    \frac{1}{M_\alpha} = \frac{1}{N_{\alpha}} + \frac{\phi^0_\alpha}{K_l}, \label{eqn:63}
\end{equation}
where $K_l$ is the fluid bulk modulus. 

\subsection{Discrete block matrix form}
The discrete counterpart to \crefrange{eqn:59}{eqn:61} is formulated using the finite-volume method (FVM) for flow, the virtual-element method (VEM) for mechanics (\citealt{Beirao2013}; \citealt{Gain2014}), and the backward Euler method for time. This hybrid numerical approach to poroelasticity was originally developed using the Matlab Reservoir Simulation Toolbox (MRST) (\citealt{Lie2012}; \citealt{Lie2019}), by \cite{Andersen2017a,Andersen2017b} for single-continuum materials and later expanded in \cite{Ashworth2019a} to dual-continuum materials. We use this hybrid framework in the current work due to its availability. However, recent works have shown the current modelling framework to be suitable for subsurface applications where complex geometrical structures can lead to irregular grids not easily handled by standard finite-element methods (\citealt{Andersen2017b}; \citealt{Coulet2019}).     

We partition our domain into disjoint elements (or cells). Accordingly, for the DC problem $\Omega^\mathcal{D} = \cup^{n_{elem}}_{j=1}\Omega^{\mathcal{D}}_j$, where $n_{elem}$ is the number of elements. Notation $\Omega^\mathcal{D}_{j}$ denotes the dual-continuum element for which there are \textit{two} pressure degrees of freedom, corresponding to each continuum.  

We define the following discrete solution spaces for the DC problem as $\mathcal{S}^h_{P_\alpha}\subset \mathcal{S}_{P_\alpha}$ and $\mathcal{S}^h_{U}\subset \mathcal{S}_{U}$. Discrete weighting spaces are given as $\mathcal{W}^h_{P_\alpha}\subset \mathcal{W}_{P_\alpha}$ and $\mathcal{W}^h_{U}\subset \mathcal{W}_{U}$. Discrete continuum pressure fields, $P^h_\alpha\in\mathcal{S}^h_{P_\alpha}$, and discrete displacement fields, $\bm{U}^h\in\mathcal{S}^h_{U}$, are given according to the following interpolation relations respectively,
\begin{align}
    P^h_\alpha = \sum^{n_{elem}}_{j=1} \mathbb{I}^j\tilde{P}^j_{\alpha} \label{eqn:64} \\
    \bm{U}^h = \sum^{n_{node}}_{b=1} \mathbb{N}^b \tilde{\bm{U}}^b \label{eqn:65}
\end{align}
where $n_{node}$ denotes the total number of vertices, and $\tilde{P}^j_{\alpha}$ and $\tilde{\bm{U}}^b$ are pressure and displacement degrees of freedom respectively, with the corresponding basis functions denoted by $\mathbb{I}^j$ and $\mathbb{N}^{b}$. 

In FVM we consider $\tilde{P}^j_{\alpha}$ to be cell-centred quantities. Notation $\mathbb{I}^j$ is then an indicator function for continuum $\alpha$ given as
\begin{equation}
    \mathbb{I}^j(\bm{X})  = 
        \begin{cases}
            1 &\text{if } \bm{X} \text{ in } \Omega^{\mathcal{D}}_j \\
            0 &\text{otherwise }  
        \end{cases}, \label{eqn:66}
\end{equation}
Further, we replace discrete pressure weight functions, $\omega^h_{\alpha}\in\mathcal{W}^h_{P_\alpha}$, by the indicator function whilst also using \cref{eqn:64} such that \crefrange{eqn:60}{eqn:61} can be interpreted as element-wise conservation statements. Using Gauss's theorem, element-wise divergence of flux volume integrals in \crefrange{eqn:60}{eqn:61}, are turned into face-wise surface integrals. In this work we use the two-point flux approximation to calculate these face-wise flux integrals (see \cite{Lie2019} for further details).

The nodal basis function matrix, $\mathbb{N}^b$, takes the identity matrix $\bm{1}$ when located at node $b$ and $\bm{0}$ at all other nodes. VEM is a Galerkin based method, thus the discrete displacement weight, $\bm{\eta}^h\in\mathcal{W}^h_{U}$, is an interpolation of the type shown in \cref{eqn:65}. However in VEM, contrary to standard finite-element methods, the bilinear form with discrete fields can never be directly calculated as basis functions are never explicitly defined. Due to basis function independence, VEM can be interpreted as a generalisation of the finite-element method to arbitrary polygonal and polyhedral meshes. Such a property is desirable for subsurface modelling, where degenerate cells and hanging nodes are encountered (\citealt{Andersen2017b}). Instead, the idea in VEM is to approximate the bilinear form, such that
\begin{equation}
    g(\bm{\eta}^h, \bm{U}^h) \approx g^h(\bm{\eta}^h, \bm{U}^h) \label{eqn:67},
\end{equation}
where $g^h(\bm{\eta}^h, \bm{U}^h)=\sum^{n_{elem}}_{j=1} g^{h}_j(\bm{\eta}^h, \bm{U}^h)$, and where details of the element-wise first-order bilinear VEM approximation, $g^{h}_j(\bm{\eta}^h, \bm{U}^h)$, can be found in \cite{Gain2014} and \cite{Andersen2017b}. Finally, as part of the VEM assembly, the constitutive relation, \cref{eqn:62}, need only be computed once, similar to a one-point quadrature finite-element scheme (\citealt{Beirao2015}).

Replacing solutions and weighting functions with their discrete counterparts, and using the time discretisation, the discrete residual equations from \crefrange{eqn:59}{eqn:60} are
\begingroup
\allowdisplaybreaks
\begin{align}
    \bm{R}^a_{U} &= g^h(\mathbb{N}^a , \bm{U}^{h,n+1}) - \int_{\Omega^{\mathcal{D}}} (\nabla\mathbb{N}^a)\cdot\bm{B}_mP^{h,n+1}_m \text{ d}V \nonumber \\
    &\phantom{=} - \int_{\Omega^{\mathcal{D}}} (\nabla\mathbb{N}^a)\cdot\bm{B}_fP^{h,n+1}_f \text{ d}V - \int_{\Gamma^{\text{T}}}\mathbb{N}^a\cdot\hat{\bm{T}} \text{ d}S \nonumber\\
    &= \bm{0}, \quad \forall a = 1,...,n_{node}, \label{eqn:68} \\
    R^i_{P_m} &= \int_{\Omega^{\mathcal{D}}_i}\bm{B}_m\cdot\Delta(\underline{\nabla}\bm{U}^{h,n+1}) + \frac{1}{M_m}\Delta P^{h,n+1}_m + \frac{1}{Q}\Delta P^{h,n+1}_f \text{ d}V \nonumber \\
    &\phantom{=} - \Delta t\int_{\partial\Omega^{\mathcal{D}}_i}\left(\frac{\bm{k}^*_m}{\mu_l}\cdot\nabla P^{h,n+1}_m\right)\cdot\bm{n} \text{ d}S \nonumber \\
    &\phantom{=}-\Delta t\int_{\Omega^{\mathcal{D}}_i}\frac{\varkappa k'}{\mu_l}(P^{h,n+1}_f - P^{h,n+1}_m) \text{ d}V \nonumber \\
    &= 0, \quad \forall i = 1,...,n_{elem},  \label{eqn:69} \\
    R^i_{P_f}  &= \int_{\Omega^{\mathcal{D}}_i} \left(\bm{B}_f\cdot\Delta(\underline{\nabla}\bm{U}^{h,n+1}) + \frac{1}{Q}\Delta P^{h,n+1}_m + \frac{1}{M_f}\Delta P^{h,n+1}_f\right) \text{ d}V \nonumber \\
    &\phantom{=}- \Delta t\int_{\partial\Omega^{\mathcal{D}}_i}\left(\frac{\bm{k}^*_f}{\mu_l}\cdot\nabla P^{h,n+1}_f\right)\cdot\bm{n} \text{ d}S \nonumber \\
    &\phantom{=}-\Delta t\int_{\Omega^{\mathcal{D}}_i}\frac{\varkappa k'}{\mu_l}(P^{h,n+1}_m - P^{h,n+1}_f) \text{ d}V \nonumber \\
    &= 0, \quad \forall i = 1,...,n_{elem}, \label{eqn:70}
\end{align}
\endgroup
where we make use of Voigt notation for tensor representation. Notation $\Delta\bm{z}^{n+1}=\bm{z}^{n+1}-\bm{z}^{n}$, where $n$ denotes the current time level. Details of the VEM calculations for the boundary and gradient terms involving $\mathbb{N}^a$ in \cref{eqn:68} can be found in \cite{Andersen2017b}. 

Even though we assume linearity in the current work, poroelastic problems are generally non-linear due to material and geometric non-linearities. To provide a general numerical framework we therefore present the discrete equations describing the DC problem following application of Newton's method. In MRST this is handled naturally using an automatic differentation framework to generate the Jacobian. We give the discrete system of equations in block matrix form as
\begin{equation}
    \begin{bmatrix}
        \bm{K} & -\bm{D}^\top_m & -\bm{D}^\top_f \\
        \bm{D}_m & \bm{F}_m & \bm{E}_m \\
        \bm{D}_f & \bm{E}_f & \bm{F}_f
    \end{bmatrix}^{(l)}
    \begin{bmatrix}
        \delta\tilde{\bm{U}} \\
        \delta\tilde{\bm{P}}_m \\
        \delta\tilde{\bm{P}}_f
    \end{bmatrix}^{n+1, (l)}
    = 
    -\begin{bmatrix}
        \bm{R}_{U} \\
        \bm{R}_{P_m} \\
        \bm{R}_{P_f} 
    \end{bmatrix}^{n+1, (l)}, \label{eqn:71}
\end{equation}
where $\bm{R}_{U}=[\bm{R}^1_{U},...,\bm{R}^{n_{node}}_U]^\top$ and $\bm{R}_{P_\alpha}=[R^1_{P_\alpha},...,R^{n_{elem}}_{P_\alpha}]^\top$. Notations $\delta$ and $l$ denote the change in solution and current iteration levels respectively. Further, $\tilde{\bm{U}}=[\tilde{\bm{U}}^1,...,\tilde{\bm{U}}^{n_{node}}]^\top$ and $\tilde{\bm{P}}_\alpha=[\tilde{P}_{\alpha}^1,...,\tilde{P}_{\alpha}^{n_{elem}}]^\top$. The individual matrices comprising the Jacobian in \cref{eqn:71} are given as 
\begin{align}
    \bm{K}_{ab} = \frac{\partial\bm{R}^a_U}{\partial\tilde{\bm{U}}^b} &= g^h(\mathbb{N}^a,\mathbb{N}^b), \label{eqn:72} \\
    \bm{D}_{ib, \alpha} = \frac{\partial R^i_{P_\alpha}}{\partial\tilde{\bm{U}}^b} &= \int_{\Omega^{\mathcal{D}}}\mathbb{I}^i(\bm{X}_i)
    \bm{B}_\alpha\cdot\nabla\mathbb{N}^b \text{ d}V, \label{eqn:73} \\
    \bm{E}_{ij,\alpha} = \frac{\partial R^i_{P_\alpha}}{\partial\tilde{P}^j_{\beta}} &= \int_{\Omega^{\mathcal{D}}}\mathbb{I}^i(\bm{X}_i) \frac{1}{Q} \mathbb{I}^j(\bm{X}_i) \text{ d}V \nonumber \\
    &\phantom{=} -\Delta t\int_{\Omega^{\mathcal{D}}} \mathbb{I}^i(\bm{X}_i) \frac{\varkappa k'}{\mu_l}\mathbb{I}^j(\bm{X}_i) \text{ d}V, \label{eqn:74} \\
    \bm{F}_{ij, \alpha} = \frac{\partial R^i_{P_\alpha}}{\partial\tilde{P}^j_{\alpha}} &= \int_{\Omega^{\mathcal{D}}}\mathbb{I}^i(\bm{X}_i) \frac{1}{M_\alpha} \mathbb{I}^j(\bm{X}_i) \text{ d}V \nonumber\\
    &\phantom{=} + \Delta t\int_{\Omega^{\mathcal{D}}} \mathbb{I}^i(\bm{X}_i) \frac{\varkappa k'}{\mu_l}\mathbb{I}^j(\bm{X}_i) \text{ d}V + \Delta t\bm{G}_{ij,\alpha}, \label{eqn:75}
\end{align}
where $\bm{X}_i$ denotes the centroid of element $\Omega^{\mathcal{D}}_i$, and $\bm{G}_{ij,\alpha}$ is the transmissibility matrix for continuum $\alpha$ arising from the two-point flux approximation (\citealt{Lie2019}). 

Finally, \cref{eqn:71} is solved using a fully coupled approach (\citealt{Lewis1998}), although extensions to sequential solution strategies for DC materials have been shown in \cite{Kim2012} and \cite{Ashworth2019a}.

\section{Numerical tests}
With the framework in-hand, we present and conduct the numerical tests used to investigate whether the macroscopic dual-continuum poroelastic model can capture global flow and deformation behaviours of a fine-scale (FS) explicit model. In doing, we review several considerations for the interpretation of the results at the scale of the DC model. 

\subsection{Test cases}
We introduce four numerical experiments to test the validity of the DC poroelastic concept. In each case we consider an idealised representation of a naturally fractured rock sample. Our idealisation comes in that we assume the fracture fabric to be periodic. To start we consider an undeformable isotropic material to understand the physics of the flow problem. We progress by introducing mechanics to the isotropic system, and then adding complexity by considering anisotropic material cases. 

In every case we consider the dimension of the domain to be $1\text{ m} \times 1 \text{ m}$. Each experiment then represents a thin 2D slice taken from a 3D sample such that, in the case of the mechanical problem, the plane-stress assumption applies. 

\subsubsection{Undeformable isotropic}
\begin{figure}[h!]
\centering
\includegraphics[scale =1.0]{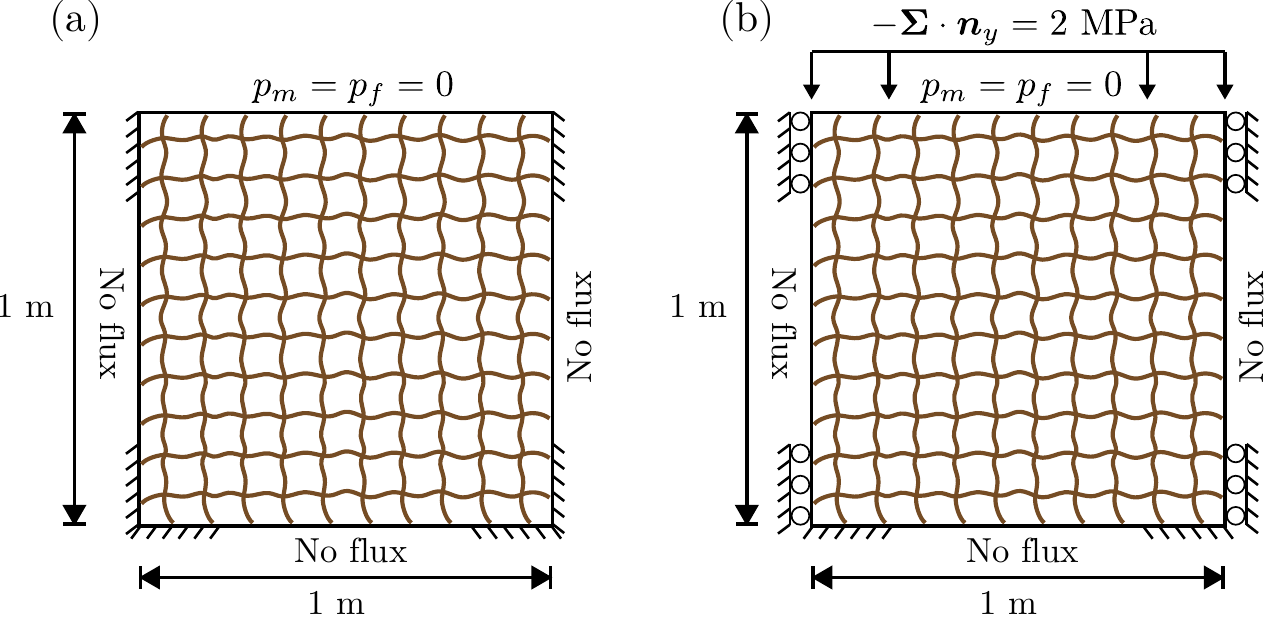}
\caption{Conceptual illustrations of the geometries and boundary conditions for (a) the undeformable isotropic problem and (b) the deformable isotropic problem. In (b) the base is fixed, whilst the left, right and top boundaries can move vertically.} \label{fig:2}
\end{figure}

For this test we study an (isotropic) undeformable matrix permeated by an isotropic undeformable fracture network. The test is setup as a uniaxial drainage problem, such that the top boundary is open to flow, $\hat{P}_m=\hat{P}_f=0$, whilst the left, right and bottom boundaries are zero flux boundaries (\cref{fig:2}a). Initial pressures for the continua are set at $P^0_m = P^0_f = 2 \text{ MPa}$. Volume fractions for matrix and fracture material are $v_m = 0.998$ and $v_f = 0.002$ respectively, given a fracture spacing, $s$, of $0.1 \text{ m}$. Local porosities for the two continua are then prescribed as $\varphi_m=0.1$ and $\varphi_f=0.9$, where the volume fractions link the global and local Lagrangian porosities so that $\phi_\alpha=v_\alpha\varphi_\alpha$. Intrinsic matrix permeability, $k_m$, is taken as $0.01 \text{ md}$, whilst individual fracture permeability, $k_f$, is calculated using the parallel plate model with a fracture aperture of $a_f \approx 1.05\times 10^{-4} \text{ m}$ (\citealt{Witherspoon1980}). The resulting permeability is $950 \text{ d}$ for each fracture \footnote{1 darcy (d) = 9.87 $\times 10^{-13} \text{ m}^{2}$.}. Fluid properties are $\rho_l = 1000 \text{ kgm}^{-3}$, $\mu_l=1 \text{ cp}$, and $K_l = 2.5 \text{ GPa}$. Upscaling individual fracture permeability to a continuum permeability for use in the DC model is done using the cubic law (\citealt{Witherspoon1980}). The resulting isotropic fracture continuum permeabitity is $k^*_{f} \approx 1000 \text{ md}$. Finally, due to the dissociation by the fracture network, the macroscopic matrix permeability is zero. 

\subsubsection{Deformable isotropic}
We now consider a deformable counterpart to the experiment described in Section 4.1.1 (\cref{fig:2}b). Accordingly, the Young's moduli for the matrix and fracture materials are $E_m = 36 \text{ GPa}$ and $E_f = 36 \text{ MPa}$ respectively. The latter is chosen for illustrative purposes as $E_f=E_m/1000$. Both continua are assigned a Poisson's ratio of $\nu = 0.2$. For an isotropic medium under the plane-stress assumption, the stiffness tensor written with Voigt notation is given as
\begin{equation}
    \mathbb{C} = 
    \begin{bmatrix}
        \dfrac{E}{1-\nu^2} & \dfrac{\nu E}{1-\nu^2} & 0 \\[1em]
                          & \dfrac{E}{1-\nu^2} & 0 \\[1em]
        sym               &                    & G
    \end{bmatrix}, \label{eqn:76}
\end{equation}
where parameter $G=E/(2(1+\nu))$ is the shear modulus. Entries for $\mathbb{C}_m$ and $\mathbb{C}_f$ can be calculated with \cref{eqn:76} and the defined intrinsic parameter values. 

For $\mathbb{C}^*$, parameters must be calculated by homogenisation. In \cite{Ashworth2019b} the author's suggest using the Hashin-Shtrikman lower bounds (\citealt{Hashin1963}), as an initial homogenisation approach for the estimation of the mechanical properties of densely fractured rock. For the bulk and shear moduli these lower bounds are quoted as (\citealt{Hashin1963}),
\begin{align}
    K^{HS^-} &= K_f + \frac{v_m}{[(K_m-K_f)^{-1} + 3v_f(3K_f + 4G_f)^{-1}]},\label{eqn:77}\\
    G^{HS^-} &= G_f \nonumber \\ 
    &+ \frac{v_m}{[(G_m-G_f)^{-1} + 6v_f(K_f+2G_f)(5G_f(3K_f + 4G_f))^{-1}]}, \label{eqn:78}
\end{align}
where $K_\alpha$ and $G_\alpha$ are the 3D bulk and shear moduli for continuum $\alpha$ respectively. 

We map between the 3D bulk modulus calculated in \cref{eqn:77} and the 2D homogenised bulk modulus under plane-stress, $K^*$, using the following relation (e.g. \citealt{Torquato2002}),   
\begin{equation}
   K^* = \frac{9K^{HS^-}G^{HS^-}}{3K^{HS^-} + 4G^{HS^-}}. \label{eqn:79}
\end{equation}
The Poissons ratio for the composite dual-continuum under plane-stress is given by
\begin{equation}
\nu^*=\frac{K^*-G^{HS^-}}{K^*+G^{HS^-}}. \label{eqn:80}
\end{equation}
Finally, the homogenised Young's modulus, $E^*$, can be recovered as
\begin{equation}
    4K^{*} = \bm{1}:\mathbb{C}^*:\bm{1} = \frac{2E^*}{(1-\nu^*)}.  \label{eqn:81} 
\end{equation}
With \crefrange{eqn:80}{eqn:81} the homogenised parameters are $\nu^*=0.2$ and $E^*=18.0 \text{ GPa}$.

We assume the matrix and fracture skeletons to be made up of the same solid material. We then assign a solid modulus, $K_s$, of 70 GPa for both continua. 

For the coupled mechanics and flow problem we consider a different method of initialisation to Section 4.1.1. Instead of assigning initial continuum pressures, we define the starting point for the experiment to be the undrained, loaded configuration, $P_m^{0+}$, $P_f^{0+}$, $\bm{U}^{0+}$. This undrained state is induced following the application of an instantaneous load on an unpressurised and undeformed domain, $P_m^0=P_f^0=0$ and $\bm{U}^0=0$. Loading is prescribed as a vertical traction of $-\bm{\Sigma}\cdot\bm{n}_y = -2 \text{ MPa}$ on the top boundary. The domain is horizontally constrained at the boundaries, but remains free to move along the vertical axis apart from at the bottom boundary where the sample is fixed. The parameters for flow are as defined in Section 4.1.1

\subsubsection{Geometry-induced anisotropy: explicit computation of $\mathbb{C}^*$}
\begin{figure}[h!]
\centering
\includegraphics[scale =1.0]{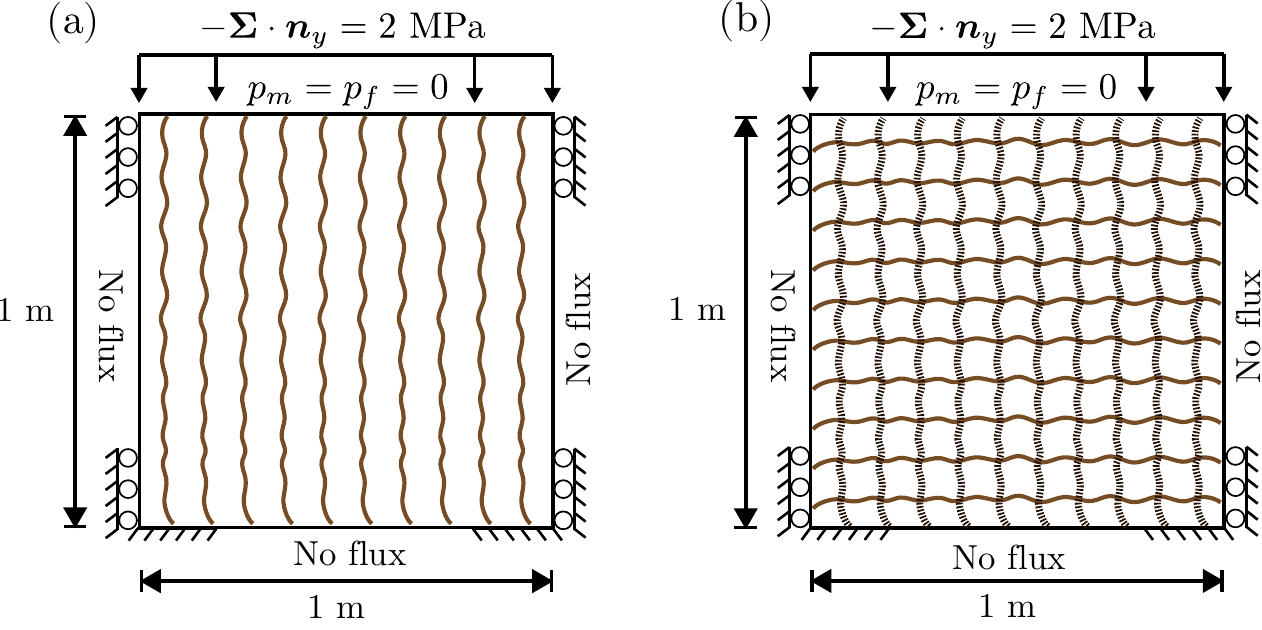}
\caption{Conceptual illustrations of the geometries and boundary conditions for (a) the anisotropic problem with one fracture set (and explicit computation of $\mathbb{C}^*$) and (b) the anisotropic problem with two fracture sets (and numerical computation of $\mathbb{C}^*$). In both cases the bases are fixed, whilst the left, right and top boundaries can move vertically.} \label{fig:3}
\end{figure}

The third experiment is concerned with an anisotropic deformable material. Anisotropy has recently been studied in poroelastic DC materials in the context of flow properties (\citealt{Zhang2019}). However, here we consider the directional dependence of \textit{both} mechanical and flow properties. Anisotropy is introduced geometrically by considering just a single vertical fracture set which is aligned with the second principal axis (\cref{fig:3}a). The 2D domain is then orthotropic. Whilst anisotropy exists at the macroscale, the intrinsic mechanical parameters remain isotropic for each continuum and are as described in Section 4.1.2. The plane-stress stiffness tensor for an orthotropic material is given by 
\begin{equation}
    \mathbb{C} = 
    \begin{bmatrix}
        \dfrac{E_1}{1-\nu_{12}\nu_{21}} & \dfrac{\nu_{21} E_1}{1-\nu_{12}\nu_{21}} & 0 \\[1em]
                          & \dfrac{E_2}{1-\nu_{12}\nu_{21}} & 0 \\[1em]
        sym               &                                 & G_{12}
    \end{bmatrix}, \label{eqn:82}
\end{equation}
Parameters of the homogenised stiffness tensor may be approximated explicitly for this geometry, using mixture theory. Accordingly, for the Young's moduli 
\begin{equation}
    E^*_1 = \left(\frac{v_m}{E_m} + \frac{v_f}{E_f}\right)^{-1}, \quad E^*_2 = v_mE_m + v_fE_f,  \label{eqn:83}
\end{equation}
where having removed a fracture set, the volume fraction of the fracture continuum is now $v_f=0.001$ (resp. $v_f=0.999$). For the homogenised Poisson's ratio, $\nu^*_{21}$, and shear modulus,  $G^*_{12}$, mixture theory gives
\begin{equation}
    \nu^*_{21} = v_m\nu_m + v_f\nu_f, \quad G^*_{12} = \left(\frac{v_m}{G_m} + \frac{v_f}{G_f}\right)^{-1}. \label{eqn:84} 
\end{equation}
The other Poisson's ratio, $\nu^*_{12}$, is readily determined by the symmetry in \cref{eqn:82} which requires $\nu_{12}E_2=\nu_{21}E_1$. From \crefrange{eqn:83}{eqn:84} and the aformentioned symmetry relation, the mechanical parameters are given as $E^*_1=18.0 \text{ GPa}$, $E^*_2=36.0 \text{ GPa}$, $\nu^*_{21}=0.200$, $\nu^*_{12}=0.100$ and $G^*_{12}=7.50 \text{ GPa}$.

The anisotropic fracture continuum leads to an anistropic permeability tensor so that permeability in the $x$ and $y$ directions are $k^*_{f,x}=0$ and $k^*_{f,y}\approx 1000 \text{ md}$ respectively. For the matrix, the macroscopic permeability is also anisotropic with $k^*_{m,x}=0$ and $k^*_{m,y}\approx 0.01 \text{ md}$. The remaining flow parameters, boundary conditions and initialisation are as described in Sections 4.1.1 and 4.1.2.

\subsubsection{Material-induced anisotropy: numerical computation of $\mathbb{C}^*$}
The final experiment is on an anisotropic material with two fracture sets aligned with each of the principal axes (\cref{fig:3}b). Anisotropy is now introduced through the fracture material, with each fracture set having different intrinsic mechanical and flow properties. These property differences are in analogy to fractures containing different amounts of infill material. To represent this conceptually within the model we assign different intrinsic porosities to the individual fracture sets. Further, we separate the intrinsic Young's moduli and permeabilities of each fracture set by two orders of magnitude. For the horizontal fracture set we assign $\varphi^\text{h}_f=0.9$, $E^\text{h}_f=3.6 \text{ MPa}$ and an intrinsic fracture permeability of $k_f = 950 \text{ d}$. For the vertical fracture set we assign $\varphi^\text{v}_f=0.4$, $E^\text{v}_f=360 \text{ MPa}$ and an intrinsic permeability of $9.5 \text{ d}$. Upscaling the fracture permeability remains trivial, with $k^*_{f,x}\approx 1000 \text{ md}$ and $k^*_{f,y}\approx 10 \text{ md}$. However, homogenistion for the parameters in the homogenised stiffness tensor now cannot be done by explicit approximation. Instead we use a deformation-driven computational homogenisation approach: We  generate unit strains for a sequence of linear displacement boundary conditions, and in doing, determine the entries of $\mathbb{C}^*$ (\citealt{Daniel1994}). Linear displacements are chosen as they produce better estimates for effective stiffness tensors for materials with a stiff matrix and weaker inclusion material (\citealt{Pecullan1999}), as is the case here. 

With the computational homogenisation approach, the mechanical parameters in $\mathbb{C}^*$ are calculated as $E^*_1=32.7 \text{ GPa}$, $E^*_2=3.40 \text{ GPa}$, $\nu^*_{21}= 0.019$, $\nu^*_{12}= 0.173$ and $G^*_{12}= 1.28 \text{ GPa}$. 

The overall volume fraction for the fracture continuum is the same as in experiment two. However, the intrinsic Lagrangian fracture porosity is now the arithmetic average of the two intrinsic fracture set porosities ($\varphi_f = 0.65$). Fluid and matrix properties remain the same as those for the other experiments. Boundary conditions and initialisation are the same as in experiments two and three. 

\subsection{Modelling considerations}
Here we review several considerations to enable the interpretation of the test results to follow. 

\subsubsection{On the REV}
Our periodic assumption of the underlying microstructure eases the requirements on our definition for an REV. In this periodic case, all the necessary geometrical and physical process information is captured within an elementary cell that is the size of the heterogeneity ($s$) (\citealt{Royer2002}; \citealt{Dormieux2006}). The separation of scales is now defined as $s \ll L$. The elementary cell definition of our REV will be useful for interpreting the discretisation choice of the DC problem. 

\subsubsection{Meshing}
\begin{figure}[h!]
\centering
\includegraphics[scale =1.0]{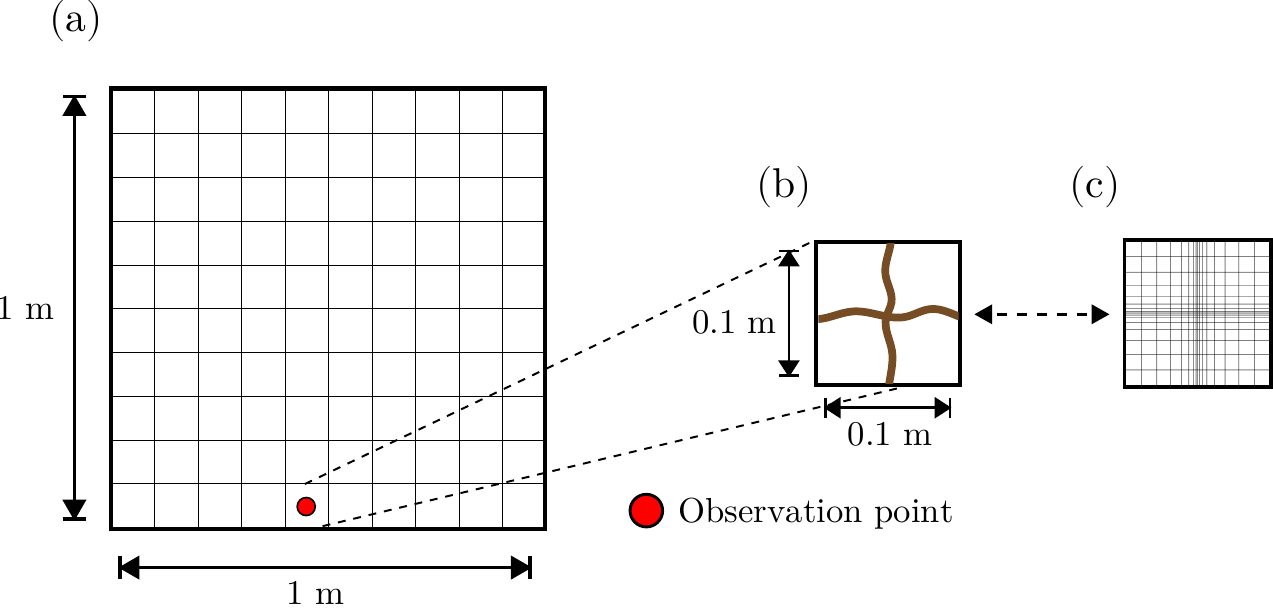}
\caption{Representations of the isotropic test problem: (a) the dual-continuum grid containing the observation point where the two modelling approaches are compared, (b) an elementary cell and (c) an equivalent fine-scale representation of the elementary cell.} \label{fig:4}
\end{figure}

For the four tests we discretise the fine-scale problem with a $200 \times 200$ Cartesian mesh, that is locally refined around the fractures (\cref{fig:4}). For the dual-continuum problem we discretise the domain using a $10 \times 10$ Cartesian mesh. In the latter case, each element then coincides with an elementary cell (in the geometrical sense) (\cref{fig:4}). 

Fine-scale and dual-continuum fields are compared at an observation point at the base of our samples. At this point we assume that our pressure solutions are sufficiently smooth, thus satisfying the physical process scale separation requirement. Further, the observation point coincides with the macroscopic material point, in this case the element centroid (\cref{fig:4}). 

\subsubsection{Quantities of interest}
At each test observation point we consider the element-wise total and continuum volumetric strains, and continuum pressures. Element-wise total and continuum volumetric strains are defined as $\text{E}^v_j=\text{tr}(\textbf{E}_j)=\Delta|\Omega^{\mathcal{D}}_j|/|\Omega^{\mathcal{D},0}_j|$, and $\text{E}^v_{j,\alpha}=\Delta|\Omega^{\mathcal{D}}_{j,\alpha}|/|\Omega^{\mathcal{D},0}_{j,\alpha}|$, respectively. We compare averaged results over the fine-scale to element level results from the dual-continuum. To enable the former, we define the following discrete continuum counterpart to \cref{eqn:2}
\begin{equation}
    \overline{z}^{\mathcal{D}}_{j,\alpha} = \frac{1}{|\Omega^{\mathcal{D}}_{j,\alpha}|}\int_{\Omega^{\mathcal{D}}_{j,\alpha}}z^{h}(\bm{x})\text{ } dV, \label{eqn:85}
\end{equation}
where $z$ is a scalar field of interest. Continuum averaged pressures and volumetric strains can then be recovered using \cref{eqn:85} with discrete microscopic fields $p^h$ or $\epsilon^{v,h}$ in place of $z^h$. Total volumetric strain is likewise obtained using \cref{eqn:85} by replacing $|\Omega^{\mathcal{D}}_{j,\alpha}|$ with $|\Omega^{\mathcal{D}}_j|$.

For the DC problem, pressures and element-wise total volumetric strain are recovered naturally from the element centroid (\citealt{Andersen2017a}). To get continuum strains, however, we must take a different approach. Starting with the matrix continuum, and comparing a volume averaged change in local porosity, \cref{eqn:4}, to the effective change in matrix porosity given by \cref{eqn:41}, such that $\text{d}\phi_m=v_m\text{d}\varphi_m$, allows us to derive the following expression for the volumetric matrix strain
\begin{equation}
    \text{E}^v_m = \bm{1}:\overline{\bm{\epsilon}}_m = \frac{1}{b_m}\left[\frac{1}{v_m}\left( B_m\bm{1}:\textbf{E} + \frac{P_m}{N_m} + \frac{P_f}{Q}\right)-\frac{P_m}{n_m}\right]. \label{eqn:86}
\end{equation}
We note the expression for $\text{E}^v_m$ in \cref{eqn:86} is only possible for an isotropic matrix as the inverse contraction map involving $\bm{b}_m$ is otherwise ill-posed. With $\text{E}^v$ and $\text{E}^v_m$ we can recover the fracture volumetric strain, $\text{E}^v_f$, for the DC model using \cref{eqn:14}.

\subsection{Results and discussion}
Here we present the results and analyses for the numerical test cases described in Section 4.1 under the modelling considerations described above. All results are given from observation points such as that shown in \cref{fig:4}.

\subsubsection{Undeformable isotropic}
\begin{figure}[h]
\centering
\setlength\figureheight{5cm}
\setlength\figurewidth{5.5cm}
\definecolor{mycolor1}{rgb}{0.57255,0.77255,0.87059}%
\definecolor{mycolor2}{rgb}{0.12941,0.40000,0.67451}%
\definecolor{mycolor3}{rgb}{0.83922,0.37647,0.30196}%
\definecolor{mycolor4}{rgb}{0.69804,0.09412,0.16863}%
\begin{tikzpicture}

\begin{axis}[%
width=0.951\figurewidth,
height=\figureheight,
at={(0\figurewidth,0\figureheight)},
scale only axis,
xmode=log,
xmin=1e-06,
xmax=10,
xminorticks=true,
xlabel style={font=\fontsize{9}{144}\selectfont\color{white!5!black}},
xlabel={Time (s)},
ymin=0,
ymax=2500000,
ylabel style={font=\fontsize{9}{144}\selectfont\color{white!5!black}},
ylabel={Pressure (Pa)},
ticklabel style={font=\fontsize{9}{144}},
axis background/.style={fill=white},
legend style={font=\fontsize{9}{144}\selectfont\color{white!5!black}, at={(1,1)}, anchor=north east, legend cell align=left, align=left, fill=none, draw=none}
]
\addplot [color=mycolor1, dashed, line width=1.5pt]
  table[row sep=crcr]{%
0	2000000\\
1e-06	1999999.99951966\\
1.74332882219999e-06	1999999.99951863\\
3.03919538231319e-06	1999999.99932488\\
5.29831690628371e-06	1999999.98696189\\
9.23670857187387e-06	1999999.64723932\\
1.61026202756094e-05	1999994.93620644\\
2.80721620394118e-05	1999957.30201714\\
4.8939009184775e-05	1999763.23054128\\
8.53167852417281e-05	1999052.91606244\\
0.000148735210729351	1997069.65054001\\
0.000259294379740467	1992629.41816083\\
0.000452035365636025	1984346.93686123\\
0.000788046281566991	1970972.61283461\\
0.00137382379588326	1951568.14669258\\
0.00239502661998749	1925442.69528259\\
0.0041753189365604	1891735.46309301\\
0.00727895384398315	1848818.986795\\
0.0126896100316792	1794029.27055595\\
0.0221221629107045	1723697.80349981\\
0.0385662042116347	1633376.98836895\\
0.0672335753649933	1518065.11587689\\
0.117210229753348	1372518.22938911\\
0.204335971785694	1192182.50935984\\
0.356224789026244	975213.454311899\\
0.621016941891562	727039.690783858\\
1.08263673387405	469141.562613203\\
1.8873918221351	244088.546931406\\
3.29034456231267	94470.2755967853\\
5.73615251044868	25203.7209404828\\
10	4356.60482378873\\
};
\addlegendentry{$\text{FS}_m$}

\addplot [color=mycolor2, line width=1.5pt]
  table[row sep=crcr]{%
0	2000000\\
1e-06	1999999.99951904\\
1.74332882219999e-06	1999999.99951892\\
3.03919538231319e-06	1999999.99951412\\
5.29831690628371e-06	1999999.99924514\\
9.23670857187387e-06	1999999.99180769\\
1.61026202756094e-05	1999999.88130394\\
2.80721620394118e-05	1999998.91370345\\
4.8939009184775e-05	1999993.41207741\\
8.53167852417281e-05	1999971.13311973\\
0.000148735210729351	1999901.69108744\\
0.000259294379740467	1999725.3734419\\
0.000452035365636025	1999344.68285026\\
0.000788046281566991	1998613.30417267\\
0.00137382379588326	1997297.70284073\\
0.00239502661998749	1994991.00911963\\
0.0041753189365604	1990974.12215017\\
0.00727895384398315	1983995.41870353\\
0.0126896100316792	1971903.37643949\\
0.0221221629107045	1951046.00898735\\
0.0385662042116347	1915350.21066053\\
0.0672335753649933	1855078.69794553\\
0.117210229753348	1755638.59748768\\
0.204335971785694	1597862.44993803\\
0.356224789026244	1363216.0029575\\
0.621016941891562	1048579.05308382\\
1.08263673387405	688459.540440422\\
1.8873918221351	360094.745917399\\
3.29034456231267	139044.740323828\\
5.73615251044868	36864.6674366276\\
10	6320.46175189616\\
};
\addlegendentry{$\text{DC}_m$}

\addplot [color=mycolor3, dashed, line width=1.0pt, mark size=2.0pt, mark=o, mark options={solid, mycolor3}]
  table[row sep=crcr]{%
0	2000000\\
1e-06	1999999.99986289\\
1.74332882219999e-06	1999999.96174656\\
3.03919538231319e-06	1999995.96732697\\
5.29831690628371e-06	1999852.28935856\\
9.23670857187387e-06	1997668.56073859\\
1.61026202756094e-05	1981408.32895629\\
2.80721620394118e-05	1914409.30615067\\
4.8939009184775e-05	1744581.65803543\\
8.53167852417281e-05	1454711.7700799\\
0.000148735210729351	1098701.08231707\\
0.000259294379740467	773074.284207851\\
0.000452035365636025	544643.015595962\\
0.000788046281566991	405585.52010552\\
0.00137382379588326	309774.989699485\\
0.00239502661998749	230732.393795511\\
0.0041753189365604	167004.965057274\\
0.00727895384398315	120243.332782385\\
0.0126896100316792	87373.2107250728\\
0.0221221629107045	64084.8529136273\\
0.0385662042116347	47112.3664986878\\
0.0672335753649933	34462.0211533112\\
0.117210229753348	24932.9886452379\\
0.204335971785694	17710.5838343\\
0.356224789026244	12215.4861587749\\
0.621016941891562	8008.10396127182\\
1.08263673387405	4768.25930935553\\
1.8873918221351	2383.82784838263\\
3.29034456231267	907.971245879314\\
5.73615251044868	240.855582561876\\
10	41.5395031028551\\
};
\addlegendentry{$\text{FS}_f$}

\addplot [color=mycolor4, line width=1.0pt, mark size=2.0pt, mark=diamond, mark options={solid, mycolor4}]
  table[row sep=crcr]{%
0	2000000\\
1e-06	1999999.99974295\\
1.74332882219999e-06	1999999.95302041\\
3.03919538231319e-06	1999996.73709443\\
5.29831690628371e-06	1999895.91112335\\
9.23670857187387e-06	1998350.09924967\\
1.61026202756094e-05	1985937.56679063\\
2.80721620394118e-05	1929367.58559625\\
4.8939009184775e-05	1769630.17077826\\
8.53167852417281e-05	1464867.75103306\\
0.000148735210729351	1043179.27525849\\
0.000259294379740467	606311.89713869\\
0.000452035365636025	273597.810968127\\
0.000788046281566991	96799.500640034\\
0.00137382379588326	34973.3305862531\\
0.00239502661998749	21399.4076391705\\
0.0041753189365604	19559.618660686\\
0.00727895384398315	19349.583184179\\
0.0126896100316792	19224.8755565099\\
0.0221221629107045	19020.8719564962\\
0.0385662042116347	18672.0484542842\\
0.0672335753649933	18083.1357354204\\
0.117210229753348	17111.6453783007\\
0.204335971785694	15570.6178201702\\
0.356224789026244	13279.4765213091\\
0.621016941891562	10209.2244662074\\
1.08263673387405	6698.11703475915\\
1.8873918221351	3500.13584201151\\
3.29034456231267	1350.05855770393\\
5.73615251044868	357.519186868294\\
10	61.2199455269859\\
};
\addlegendentry{$\text{DC}_f$}

\end{axis}
\end{tikzpicture}%
\caption{Matrix and fracture continuum element averaged pressure evolutions for the undeformable isotropic test case. `$\text{FS}_\alpha$' and `$\text{DC}_\alpha$' denote quantities related to fine-scale and dual-continuum models for continuum $\alpha$ respectively.} \label{fig:5}
\end{figure}
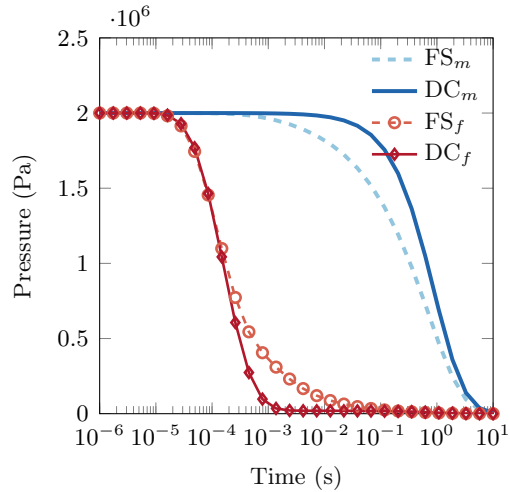

\cref{fig:5} shows the element averaged pressure evolutions from both fine-scale explicit and dual-continuum simulations for the undeformable isotropic material case. Both models show a rapid decrease in fracture pressure within the first millisecond followed by a delayed pressure response in the matrix. These general patterns can be attributed to the contrast in continuum permeabilities. Whilst both models show general decreasing trends, the FS fracture pressure decrease begins to smooth out at lower pressures. Further, the onset of fine-scale matrix pressure diffusion happens earlier. When matrix pressure diffusion does occur in the DC model, the process occurs more rapidly (indicated by a steeper gradient) than in the fine-scale case.

The disparities in matrix and fracture pressure diffusion between the two modelling approaches arise from the first-order transfer term, \cref{eqn:46}, used by the dual-continuum model. In using a linear mass transfer model one implicitly places a pseudosteady state diffusion assumption on the communication between matrix and fracture continua. As a result, transient matrix drainage effects are neglected by the DC approach. Neglecting transient effects leads to the delay in DC matrix pressure diffusion we see in \cref{fig:5}, and the loss of pressure support in the fractures. 

Shortcomings of using simplified transfer concepts have been well documented in literature (e.g. \citealt{Berkowitz1988}; \citealt{Lemonnier2010}). Previous works have thus sought to improve on the linear inter-continuum flow coupling term by including transient effects (\citealt{Zimmerman1993}; \citealt{Sarma2004}; \citealt{March2016}; \citealt{Zhou2017}). However, in the current work, we acknowledge the shortcomings of the transfer term used herein, with the focus being on understanding the coupled poroelastic behaviour. 

\subsubsection{Deformable isotropic}
\begin{figure}[h]
\centering
\begin{minipage}[b]{0.5\textwidth}
\centering
\setlength\figureheight{5cm}
\setlength\figurewidth{5.5cm}
\definecolor{mycolor1}{rgb}{0.57255,0.77255,0.87059}%
\definecolor{mycolor2}{rgb}{0.12941,0.40000,0.67451}%
\definecolor{mycolor3}{rgb}{0.83922,0.37647,0.30196}%
\definecolor{mycolor4}{rgb}{0.69804,0.09412,0.16863}%
\begin{tikzpicture}

\begin{axis}[%
width=0.951\figurewidth,
height=\figureheight,
at={(0\figurewidth,0\figureheight)},
scale only axis,
xmode=log,
xmin=1e-06,
xmax=10,
xminorticks=true,
xlabel style={font=\fontsize{9}{144}\selectfont\color{white!5!black}},
xlabel={Time (s)},
ymin=0,
ymax=2000000,
ylabel style={font=\fontsize{9}{144}\selectfont\color{white!5!black}},
ylabel={Pressure (Pa)},
ticklabel style={font=\fontsize{9}{144}},
axis background/.style={fill=white},
legend style={font=\fontsize{9}{144}\selectfont\color{white!5!black}, at={(1,1)}, anchor=north east, legend cell align=left, align=left, fill=none, draw=none}
]
\addplot [color=mycolor1, dashed, line width=1.5pt]
  table[row sep=crcr]{%
0	609042.229201229\\
1e-06	621484.101384076\\
1.74332882219999e-06	632411.384170773\\
3.03919538231319e-06	644916.375701581\\
5.29831690628371e-06	659756.951963503\\
9.23670857187387e-06	675571.854479361\\
1.61026202756094e-05	689093.872813486\\
2.80721620394118e-05	697651.853635857\\
4.8939009184775e-05	701487.08603328\\
8.53167852417281e-05	702867.161993433\\
0.000148735210729351	703745.276457656\\
0.000259294379740467	705066.93840306\\
0.000452035365636025	707197.569380982\\
0.000788046281566991	709673.534709239\\
0.00137382379588326	709765.225425447\\
0.00239502661998749	701739.118223398\\
0.0041753189365604	679578.411753951\\
0.00727895384398315	641597.89561101\\
0.0126896100316792	592776.678145543\\
0.0221221629107045	542594.264633456\\
0.0385662042116347	498269.433396761\\
0.0672335753649933	459424.275656846\\
0.117210229753348	420633.001008655\\
0.204335971785694	376755.276050612\\
0.356224789026244	324679.565821775\\
0.621016941891562	263084.432362844\\
1.08263673387405	193287.0499162\\
1.8873918221351	121808.292805768\\
3.29034456231267	61182.401927001\\
5.73615251044868	22596.6425681114\\
10	5697.35366757783\\
};
\addlegendentry{$\text{FS}_m$}

\addplot [color=mycolor2, line width=1.5pt]
  table[row sep=crcr]{%
0	708302.582616037\\
1e-06	708302.807963895\\
1.74332882219999e-06	708303.200818796\\
3.03919538231319e-06	708303.885692507\\
5.29831690628371e-06	708305.079647843\\
9.23670857187387e-06	708307.16109017\\
1.61026202756094e-05	708310.789684722\\
2.80721620394118e-05	708317.115381884\\
4.8939009184775e-05	708328.142411427\\
8.53167852417281e-05	708347.341898998\\
0.000148735210729351	708379.909173058\\
0.000259294379740467	708417.976816269\\
0.000452035365636025	708283.368908254\\
0.000788046281566991	706927.032604866\\
0.00137382379588326	701306.263634653\\
0.00239502661998749	687047.227354123\\
0.0041753189365604	662120.196423723\\
0.00727895384398315	629868.645300596\\
0.0126896100316792	598005.783377408\\
0.0221221629107045	573609.768017157\\
0.0385662042116347	557329.007040761\\
0.0672335753649933	543333.971636092\\
0.117210229753348	524334.484197886\\
0.204335971785694	494022.764382637\\
0.356224789026244	446378.213425177\\
0.621016941891562	376354.548148017\\
1.08263673387405	284178.661000847\\
1.8873918221351	181529.579404314\\
3.29034456231267	91413.2039273691\\
5.73615251044868	33624.4062366879\\
10	8413.84576637906\\
};
\addlegendentry{$\text{DC}_m$}

\addplot [color=mycolor3, dashed, line width=1.0pt, mark size=2.0pt, mark=o, mark options={solid, mycolor3}]
  table[row sep=crcr]{%
0	1349661.74684495\\
1e-06	1377077.77617841\\
1.74332882219999e-06	1403669.07358882\\
3.03919538231319e-06	1438720.85364718\\
5.29831690628371e-06	1482187.05322186\\
9.23670857187387e-06	1527852.67272111\\
1.61026202756094e-05	1565310.49797461\\
2.80721620394118e-05	1587118.32006659\\
4.8939009184775e-05	1594435.95100861\\
8.53167852417281e-05	1593557.55371472\\
0.000148735210729351	1589355.96646717\\
0.000259294379740467	1583560.33516697\\
0.000452035365636025	1575229.27055338\\
0.000788046281566991	1556581.83389357\\
0.00137382379588326	1502276.68563422\\
0.00239502661998749	1369909.22196446\\
0.0041753189365604	1131894.94016138\\
0.00727895384398315	811295.096354826\\
0.0126896100316792	481515.831421778\\
0.0221221629107045	226587.047184691\\
0.0385662042116347	84177.6214397699\\
0.0672335753649933	28049.0198153368\\
0.117210229753348	11453.8603476484\\
0.204335971785694	6527.92107718027\\
0.356224789026244	4308.10537929601\\
0.621016941891562	2890.51828969811\\
1.08263673387405	1867.24776758396\\
1.8873918221351	1094.97615634155\\
3.29034456231267	534.24218513269\\
5.73615251044868	196.292596399922\\
10	49.7428468469221\\
};
\addlegendentry{$\text{FS}_f$}

\addplot [color=mycolor4, line width=1.0pt, mark size=2.0pt, mark=diamond, mark options={solid, mycolor4}]
  table[row sep=crcr]{%
0	1588345.79196234\\
1e-06	1588345.3572314\\
1.74332882219999e-06	1588344.5993534\\
3.03919538231319e-06	1588343.27812585\\
5.29831690628371e-06	1588340.97480094\\
9.23670857187387e-06	1588336.95937603\\
1.61026202756094e-05	1588329.95925435\\
2.80721620394118e-05	1588317.75596393\\
4.8939009184775e-05	1588296.47878856\\
8.53167852417281e-05	1588259.14075656\\
0.000148735210729351	1588184.38942917\\
0.000259294379740467	1587853.37013672\\
0.000452035365636025	1585120.67819629\\
0.000788046281566991	1568343.89603347\\
0.00137382379588326	1504368.74498076\\
0.00239502661998749	1346034.3719111\\
0.0041753189365604	1072631.8085449\\
0.00727895384398315	723867.533711008\\
0.0126896100316792	390061.27012066\\
0.0221221629107045	157776.240012486\\
0.0385662042116347	46742.0109584962\\
0.0672335753649933	12245.3262470179\\
0.117210229753348	5465.59234794956\\
0.204335971785694	4431.27652984908\\
0.356224789026244	3956.27214443425\\
0.621016941891562	3333.09787467663\\
1.08263673387405	2515.79839328037\\
1.8873918221351	1606.20631000646\\
3.29034456231267	808.273145330845\\
5.73615251044868	297.052618212219\\
10	74.2593919689208\\
};
\addlegendentry{$\text{DC}_f$}

\end{axis}
\end{tikzpicture}%
\subcaption{Pressure}\label{fig:6a}
\end{minipage}%
\begin{minipage}[b]{0.5\textwidth}
\centering
\setlength\figureheight{5cm}
\setlength\figurewidth{5.5cm}
\definecolor{mycolor1}{rgb}{0.69804,0.09412,0.16863}%
\definecolor{mycolor2}{rgb}{0.12941,0.40000,0.67451}%
\begin{tikzpicture}

\begin{axis}[%
width=0.951\figurewidth,
height=\figureheight,
at={(0\figurewidth,0\figureheight)},
scale only axis,
xmode=log,
xmin=1e-06,
xmax=10,
xminorticks=true,
xlabel style={font=\fontsize{9}{144}\selectfont\color{white!5!black}},
xlabel={Time (s)},
ymin=-12e-05,
ymax=-4e-05,
ylabel style={font=\fontsize{9}{144}\selectfont\color{white!5!black}},
ylabel={Volumetric Strain},
ticklabel style={font=\fontsize{9}{144}},
axis background/.style={fill=white},
legend style={font=\fontsize{9}{144}\selectfont\color{white!5!black}, at={(1,1)}, anchor=north east, legend cell align=left, align=left, fill=none, draw=none}
]
\addplot [color=mycolor1, dashed, line width=1.5pt]
  table[row sep=crcr]{%
0	-4.25301633256629e-05\\
1e-06	-4.33982496140172e-05\\
1.74332882219999e-06	-4.41623600409129e-05\\
3.03919538231319e-06	-4.5039960507467e-05\\
5.29831690628371e-06	-4.60844007155811e-05\\
9.23670857187387e-06	-4.7206386537297e-05\\
1.61026202756094e-05	-4.8187058973715e-05\\
2.80721620394118e-05	-4.88437167464804e-05\\
4.8939009184775e-05	-4.91906462840107e-05\\
8.53167852417281e-05	-4.93803083201855e-05\\
0.000148735210729351	-4.95356038916058e-05\\
0.000259294379740467	-4.97025458455076e-05\\
0.000452035365636025	-4.99258441002274e-05\\
0.000788046281566991	-5.04524145891499e-05\\
0.00137382379588326	-5.21094819464809e-05\\
0.00239502661998749	-5.63122248854748e-05\\
0.0041753189365604	-6.40272664600745e-05\\
0.00727895384398315	-7.45746941985955e-05\\
0.0126896100316792	-8.55986597631916e-05\\
0.0221221629107045	-9.43381583254593e-05\\
0.0385662042116347	-9.95044769533198e-05\\
0.0672335753649933	-0.000101914054188718\\
0.117210229753348	-0.000103104292971628\\
0.204335971785694	-0.000104025249615328\\
0.356224789026244	-0.000105008826344672\\
0.621016941891562	-0.000106138017323333\\
1.08263673387405	-0.000107402684522128\\
1.8873918221351	-0.000108691383859497\\
3.29034456231267	-0.000109782280767746\\
5.73615251044868	-0.000110476109113155\\
10	-0.000110779908421667\\
};
\addlegendentry{FS}

\addplot [color=mycolor2, line width=1.5pt]
  table[row sep=crcr]{%
0	-5.14888849947326e-05\\
1e-06	-5.14888925333842e-05\\
1.74332882219999e-06	-5.14889056757154e-05\\
3.03919538231319e-06	-5.14889285870681e-05\\
5.29831690628371e-06	-5.1488968528931e-05\\
9.23670857187387e-06	-5.14890381602499e-05\\
1.61026202756094e-05	-5.14891595490759e-05\\
2.80721620394118e-05	-5.14893711662099e-05\\
4.8939009184775e-05	-5.14897401733727e-05\\
8.53167852417281e-05	-5.14903904547172e-05\\
0.000148735210729351	-5.14917984480823e-05\\
0.000259294379740467	-5.14999522965634e-05\\
0.000452035365636025	-5.15753797495757e-05\\
0.000788046281566991	-5.20480303631708e-05\\
0.00137382379588326	-5.38585058880133e-05\\
0.00239502661998749	-5.83456068065747e-05\\
0.0041753189365604	-6.60991910399256e-05\\
0.00727895384398315	-7.59982010939116e-05\\
0.0126896100316792	-8.54906347809447e-05\\
0.0221221629107045	-9.2136311346402e-05\\
0.0385662042116347	-9.53965043308026e-05\\
0.0672335753649933	-9.65709501353082e-05\\
0.117210229753348	-9.70955023381054e-05\\
0.204335971785694	-9.76710785928175e-05\\
0.356224789026244	-9.85450422719188e-05\\
0.621016941891562	-9.98275134179669e-05\\
1.08263673387405	-0.000101515618004828\\
1.8873918221351	-0.00010339551291282\\
3.29034456231267	-0.000105045870185405\\
5.73615251044868	-0.000106104179879799\\
10	-0.000106565864997555\\
};
\addlegendentry{DC}

\end{axis}
\end{tikzpicture}%
\subcaption{Volumetric strain, $\text{E}^v$}\label{fig:6b}
\end{minipage}
\caption{Matrix and fracture continuum element averaged pressure (a) and total element volumetric strain (b) evolutions for the deformable isotropic test case. `$\text{FS}_\alpha$' and `$\text{DC}_\alpha$' denote quantities related to fine-scale and dual-continuum models for continuum $\alpha$ respectively.} \label{fig:6}
\end{figure}
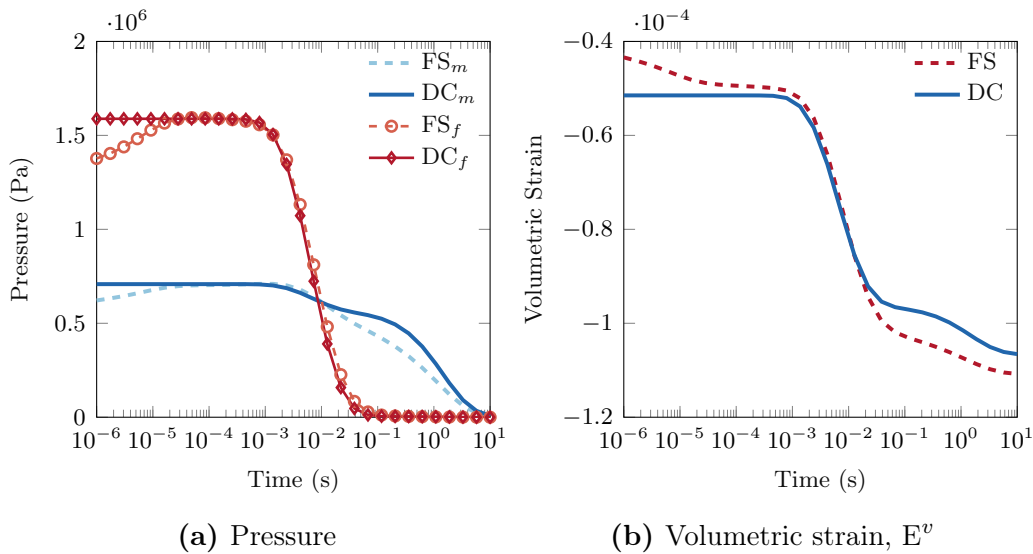

Pressure and total element volumetric strain results for the deformable isotropic case are shown in \cref{fig:6}. In \cref{fig:6a} both modelling approaches predict higher induced initial pressures in the fracture than in the matrix. Further, both approaches show rapid decreases in fracture pressure and gradual decreases in matrix pressure. In \cref{fig:6b} the FS and DC models show increasing volumetric strain evolution behaviours. However, for both pressure and strain, specific differences of the variable fields between the two modelling approaches may be observed at both early and late times. The disparity in late-time matrix pressure evolution (\cref{fig:6a}), is again due to the first-order inter-continuum transfer model (see Section 4.3.1). Over the same late-time period, we also observe a difference in volumetric strain (\cref{fig:6b}).

Of more interest in \cref{fig:6} are the early-time results for continuum pressures. In \cref{fig:6a} both fine-scale matrix and fracture pressures exhibit non-monotonic behaviour, known as the Mandel-Cryer effect. These pressure rises are not seen in the dual-continuum pressure responses. A similar observation was also made in the work of \cite{Zhang2019}, albeit for a different problem.  

\begin{figure}[h]
\centering
\begin{minipage}[b]{0.5\textwidth}
\centering
\setlength\figureheight{5cm}
\setlength\figurewidth{5.5cm}
\definecolor{mycolor1}{rgb}{0.69804,0.09412,0.16863}%
\definecolor{mycolor2}{rgb}{0.12941,0.40000,0.67451}%
\begin{tikzpicture}

\begin{axis}[%
width=0.951\figurewidth,
height=\figureheight,
at={(0\figurewidth,0\figureheight)},
scale only axis,
xmode=log,
xmin=1e-06,
xmax=10,
xminorticks=true,
xlabel style={font=\fontsize{9}{144}\selectfont\color{white!5!black}},
xlabel={Time (s)},
ymin=-5.5e-05,
ymax=-3.5e-05,
ylabel style={font=\fontsize{9}{144}\selectfont\color{white!5!black}},
ylabel={Volumetric Strain},
ticklabel style={font=\fontsize{9}{144}},
axis background/.style={fill=white},
legend style={font=\fontsize{9}{144}\selectfont\color{white!5!black}, at={(1,1)}, anchor=north east, legend cell align=left, align=left, fill=none, draw=none}
]
\addplot [color=mycolor1, dashed, line width=1.5pt]
  table[row sep=crcr]{%
0	-4.15847294650428e-05\\
1e-06	-4.24334828363502e-05\\
1.74332882219999e-06	-4.31783132384833e-05\\
3.03919538231319e-06	-4.40299170477809e-05\\
5.29831690628371e-06	-4.50392742221831e-05\\
9.23670857187387e-06	-4.61120963798349e-05\\
1.61026202756094e-05	-4.70230020707086e-05\\
2.80721620394118e-05	-4.75857298320055e-05\\
4.8939009184775e-05	-4.78102613707918e-05\\
8.53167852417281e-05	-4.78406819264263e-05\\
0.000148735210729351	-4.77931194591578e-05\\
0.000259294379740467	-4.77060850651322e-05\\
0.000452035365636025	-4.75693875771349e-05\\
0.000788046281566991	-4.73160036948201e-05\\
0.00137382379588326	-4.67604744950582e-05\\
0.00239502661998749	-4.56274911702244e-05\\
0.0041753189365604	-4.37996115497145e-05\\
0.00727895384398315	-4.1552571669331e-05\\
0.0126896100316792	-3.94930558020013e-05\\
0.0221221629107045	-3.8218886283541e-05\\
0.0385662042116347	-3.79212704901006e-05\\
0.0672335753649933	-3.83449541556026e-05\\
0.117210229753348	-3.91577277710252e-05\\
0.204335971785694	-4.02083297692189e-05\\
0.356224789026244	-4.14816382830818e-05\\
0.621016941891562	-4.29872882535981e-05\\
1.08263673387405	-4.46867264958434e-05\\
1.8873918221351	-4.64214782693153e-05\\
3.29034456231267	-4.78906470004992e-05\\
5.73615251044868	-4.88254767235298e-05\\
10	-4.92350337701423e-05\\
};
\addlegendentry{FS}

\addplot [color=mycolor2, line width=1.5pt]
  table[row sep=crcr]{%
0	-5.03741041471488e-05\\
1e-06	-5.0374094856144e-05\\
1.74332882219999e-06	-5.03740786588886e-05\\
3.03919538231319e-06	-5.03740504218108e-05\\
5.29831690628371e-06	-5.0374001195495e-05\\
9.23670857187387e-06	-5.03739153784341e-05\\
1.61026202756094e-05	-5.03737657728837e-05\\
2.80721620394118e-05	-5.03735049664854e-05\\
4.8939009184775e-05	-5.03730502879839e-05\\
8.53167852417281e-05	-5.03722560481096e-05\\
0.000148735210729351	-5.03708076200923e-05\\
0.000259294379740467	-5.0366954962261e-05\\
0.000452035365636025	-5.03459810352542e-05\\
0.000788046281566991	-5.02300034963339e-05\\
0.00137382379588326	-4.97985701113624e-05\\
0.00239502661998749	-4.87391974340958e-05\\
0.0041753189365604	-4.69173027593817e-05\\
0.00727895384398315	-4.46041580826587e-05\\
0.0126896100316792	-4.24142190458498e-05\\
0.0221221629107045	-4.09439682461046e-05\\
0.0385662042116347	-4.03526384012406e-05\\
0.0672335753649933	-4.03845669272245e-05\\
0.117210229753348	-4.07829300909646e-05\\
0.204335971785694	-4.15051834262781e-05\\
0.356224789026244	-4.26506353095796e-05\\
0.621016941891562	-4.43347815027489e-05\\
1.08263673387405	-4.65517397562685e-05\\
1.8873918221351	-4.90205979555002e-05\\
3.29034456231267	-5.11880318856225e-05\\
5.73615251044868	-5.25779432617996e-05\\
10	-5.3184298874468e-05\\
};
\addlegendentry{DC}

\end{axis}
\end{tikzpicture}%
\subcaption{Matrix vol. strain, $\text{E}^v_m$}\label{fig:7a}
\end{minipage}%
\begin{minipage}[b]{0.5\textwidth}
\centering
\setlength\figureheight{5cm}
\setlength\figurewidth{5.5cm}
\definecolor{mycolor1}{rgb}{0.69804,0.09412,0.16863}%
\definecolor{mycolor2}{rgb}{0.12941,0.40000,0.67451}%
\begin{tikzpicture}

\begin{axis}[%
width=0.951\figurewidth,
height=\figureheight,
at={(0\figurewidth,0\figureheight)},
scale only axis,
xmode=log,
xmin=1e-06,
xmax=10,
xminorticks=true,
xlabel style={font=\fontsize{9}{144}\selectfont\color{white!5!black}},
xlabel={Time (s)},
ymin=-0.035,
ymax=0,
ylabel style={font=\fontsize{9}{144}\selectfont\color{white!5!black}},
ylabel={Volumetric Strain},
ticklabel style={font=\fontsize{9}{144}},
axis background/.style={fill=white},
legend style={font=\fontsize{9}{144}\selectfont\color{white!5!black}, at={(1,1)}, anchor=north east, legend cell align=left, align=left, fill=none, draw=none}
]
\addplot [color=mycolor1, dashed, line width=1.5pt]
  table[row sep=crcr]{%
0	-0.000487780581483668\\
1e-06	-0.000497753471141007\\
1.74332882219999e-06	-0.00050759747540847\\
3.03919538231319e-06	-0.000520718156006027\\
5.29831690628371e-06	-0.000538284889397326\\
9.23670857187387e-06	-0.000562560416566036\\
1.61026202756094e-05	-0.000596397615430211\\
2.80721620394118e-05	-0.000641290447329924\\
4.8939009184775e-05	-0.000699280499662635\\
8.53167852417281e-05	-0.000774464657614352\\
0.000148735210729351	-0.00087015559064818\\
0.000259294379740467	-0.000989932248127581\\
0.000452035365636025	-0.00115969490990238\\
0.000788046281566991	-0.0015275396233123\\
0.00137382379588326	-0.0025712151578466\\
0.00239502661998749	-0.00508826882468591\\
0.0041753189365604	-0.00959020586269355\\
0.00727895384398315	-0.0156262853712164\\
0.0126896100316792	-0.0217989519658133\\
0.0221221629107045	-0.0265236108807958\\
0.0385662042116347	-0.0291020075301805\\
0.0672335753649933	-0.0300396702004603\\
0.117210229753348	-0.0302186269977134\\
0.204335971785694	-0.0301584917020849\\
0.356224789026244	-0.0300230266031489\\
0.621016941891562	-0.0298468630923063\\
1.08263673387405	-0.0296433730992771\\
1.8873918221351	-0.0294345941863068\\
3.29034456231267	-0.0292575386317767\\
5.73615251044868	-0.0291447332841011\\
10	-0.029095230662174\\
};
\addlegendentry{FS}

\addplot [color=mycolor2, line width=1.5pt]
  table[row sep=crcr]{%
0	-0.00060776452793904\\
1e-06	-0.000607772933476214\\
1.74332882219999e-06	-0.000607787587072276\\
3.03919538231319e-06	-0.000607813133050434\\
5.29831690628371e-06	-0.000607857667913522\\
9.23670857187387e-06	-0.000607935306286287\\
1.61026202756094e-05	-0.000608070653868994\\
2.80721620394118e-05	-0.000608306604828733\\
4.8939009184775e-05	-0.000608717992982385\\
8.53167852417281e-05	-0.000609439459351918\\
0.000148735210729351	-0.000610866221615069\\
0.000259294379740467	-0.000616865622113476\\
0.000452035365636025	-0.000665045338195974\\
0.000788046281566991	-0.000959243436914779\\
0.00137382379588326	-0.00207976645843681\\
0.00239502661998749	-0.00485194388367356\\
0.0041753189365604	-0.00963786144303129\\
0.00727895384398315	-0.0157416256637091\\
0.0126896100316792	-0.0215806220865933\\
0.0221221629107045	-0.0256371155183948\\
0.0385662042116347	-0.0275622856031822\\
0.0672335753649933	-0.028133576170969\\
0.117210229753348	-0.0281970690536614\\
0.204335971785694	-0.0281244527666959\\
0.356224789026244	-0.0279898541164791\\
0.621016941891562	-0.0277907007391117\\
1.08263673387405	-0.027528490864036\\
1.8873918221351	-0.0272364780766156\\
3.29034456231267	-0.0269801071817771\\
5.73615251044868	-0.0268156962522614\\
10	-0.0267439673604177\\
};
\addlegendentry{DC}

\end{axis}
\end{tikzpicture}%
\subcaption{Fracture vol. strain, $\text{E}^v_f$}\label{fig:7b}
\end{minipage}
\caption{Matrix (a) and fracture (b) volumetric strain evolutions for the deformable isotropic test case. `$\text{FS}$' and `$\text{DC}$' denote quantities related to fine-scale and dual-continuum models.} \label{fig:7}
\end{figure}
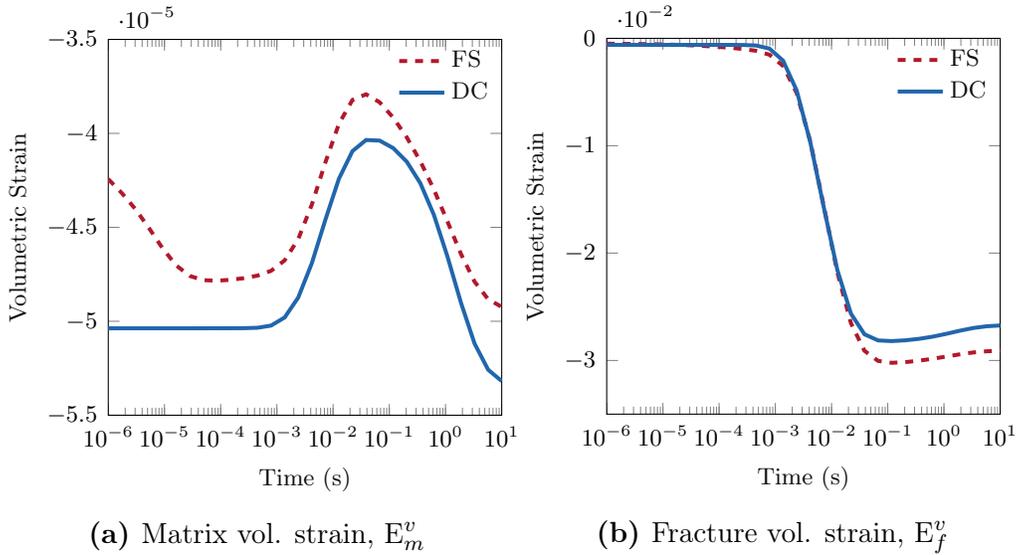

\cref{fig:7} shows individual continuum volumetric strain evolutions. In \cref{fig:7b} both modelling approaches show similar increasing strain behaviour with time. However in \cref{fig:7a}, we observe the FS matrix strain shows early-time non-monotonic behaviour, contrary to DC matrix strain. 

The early non-monotonic differences in pressure and matrix strain between the two modelling approaches result from the underlying pressure assumption made for the DC model. In the derivation of the constitutive model in Section 2.2 we assumed a local equilibrium of pressure within each continuum over an REV. The induced response predicted in by the DC model is thus for a system in mechanical and hydrostatic equilibrium. Instead, the fine-scale model makes no such pressure assumption. To understand the specific impacts of the latter it is interesting to look at the local flow and deformation behaviours shown by the fine-scale. 

\begin{figure}[h!]
\centering
\includegraphics[scale =1.0]{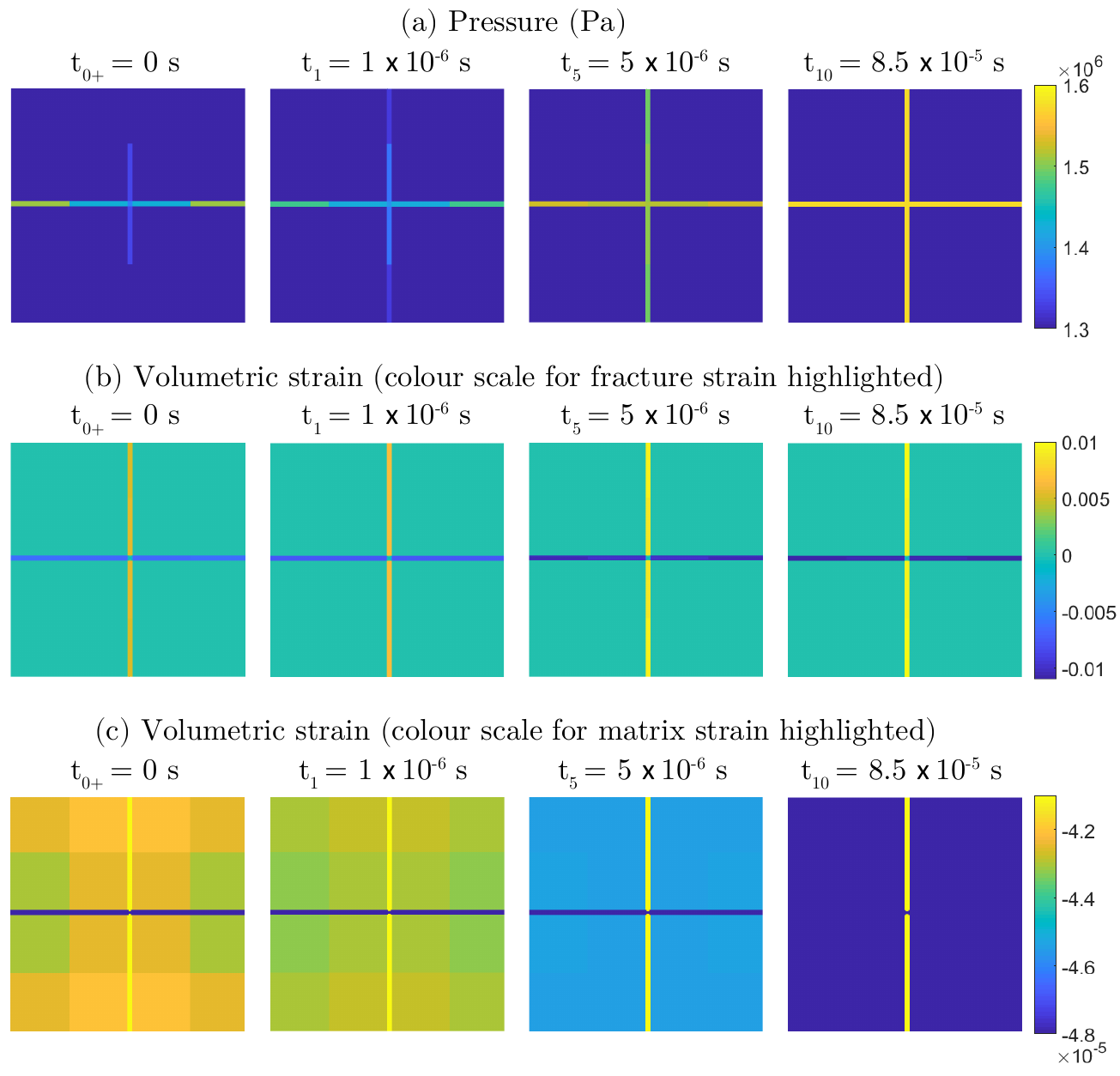}
\caption{Pressure (a) and volumetric strain (highlighted for fractures and matrix, (b) and (c) respectively) fields at different time levels, $t_i$, for the FS representation of the deformable isotropic material. Each field plot is $5 \text{ mm} \times 5 \text{ mm}$ and is located at the observation point. Subscript $0+$ denotes the time level corresponding to the undrained, loaded configuration.} \label{fig:8}
\end{figure}

\cref{fig:8} shows the FS pressure and volumetric strain responses within the first 100 microseconds. At $t_{0+}$ in \cref{fig:8}a, we observe pressure in the horizontal fracture is higher than the vertical fracture. This disequilibrium is concurrent with the negative and positive fracture strains for the horizontal and vertical fractures respectively (\cref{fig:8}b). From $t_{0+}$ to $t_{1}$, \cref{fig:8}a shows, away from the fracture intersection, horizontal fracture pressure drops slightly. However, vertical fracture pressure increases. These pressure changes occur with further contraction and expansion respectively (\cref{fig:8}b). Over the same time period matrix strain increases (\cref{fig:8}c). From $t_{1}$ to $t_{5}$ the pressure in both fractures is increasing (\cref{fig:8}a), with matrix and fracture deformations following the same evolution paths described previously. Finally, at $t_{10}$ the fractures reach a pressure equilibrium (\cref{fig:8}a).

We can now explain the early-time non-monotonic behaviours in \cref{fig:6a} and \cref{fig:7a} with the description of the local processes shown in \cref{fig:8}. Following $t_{0+}$, intra-fracture flow is driven by the pressure disequilibrium between the horizontal and vertical fractures. Between $t_{0+}$ and $t_{1}$ horizontal fracture contraction occurs primarily due to the dissipation of the fluid pressure support. Vertical fracture expansion follows due to poroelastic coupling to accommodate incoming fluid from the horizontal fracture. As the vertical fracture expands it forces the contraction of the matrix, and thus the increasing matrix strain shown in \cref{fig:7a} and \cref{fig:8}. After $t_{1}$, deformation drives the horizontal fracture pressure increase due to fluid compressibility. The overall fracture continuum pressure increases (\cref{fig:6a}), with strain generating pressure in the horizontal fracture, whilst the pressure change associated with vertical fracture expansion slows. The latter occurs due to the low matrix permeability which prohibits dissipation of excess matrix pressure, until later times. As a result, the undrained matrix stiffness increases with its progressive contraction, slowing vertical fracture expansion until a mechanical equilibrium is reached. The overall fracture continuum pressure rise finally stops when the fractures have reached mechanical equilibrium with the matrix and an internal hydrostatic equilibrium. 

The local processes shown by the FS model are not captured by the DC model due to the underlying homogenisation assumptions made in the latter. However, \crefrange{fig:6}{fig:7} do show that, aside from the local equilibration processes, the DC model can capture the global poroelastic behaviours of the FS model.

\subsubsection{Geometry-induced anisotropy: explicit computation of $\mathbb{C}^*$}
\begin{figure}[h]
\centering
\begin{minipage}[b]{0.5\textwidth}
\centering
\setlength\figureheight{5cm}
\setlength\figurewidth{5.5cm}
\definecolor{mycolor1}{rgb}{0.57255,0.77255,0.87059}%
\definecolor{mycolor2}{rgb}{0.12941,0.40000,0.67451}%
\definecolor{mycolor3}{rgb}{0.83922,0.37647,0.30196}%
\definecolor{mycolor4}{rgb}{0.69804,0.09412,0.16863}%
\begin{tikzpicture}

\begin{axis}[%
width=0.951\figurewidth,
height=\figureheight,
at={(0\figurewidth,0\figureheight)},
scale only axis,
xmode=log,
xmin=1e-05,
xmax=100,
xminorticks=true,
xlabel style={font=\fontsize{9}{144}\selectfont\color{white!5!black}},
xlabel={Time (s)},
ymin=0,
ymax=1000000,
ylabel style={font=\fontsize{9}{144}\selectfont\color{white!5!black}},
ylabel={Pressure (Pa)},
ticklabel style={font=\fontsize{9}{144}},
axis background/.style={fill=white},
legend style={font=\fontsize{9}{144}\selectfont\color{white!5!black}, at={(1,1)}, anchor=north east, legend cell align=left, align=left, fill=none, draw=none}
]
\addplot [color=mycolor1, dashed, line width=1.5pt]
  table[row sep=crcr]{%
0	609008.811361176\\
1e-05	609000.097067049\\
1.74332882219999e-05	609027.464828029\\
3.03919538231319e-05	609068.700407039\\
5.29831690628371e-05	609115.792543165\\
9.23670857187387e-05	609163.529566332\\
0.000161026202756094	609139.664315593\\
0.000280721620394118	608541.485992161\\
0.000489390091847749	605787.167120873\\
0.000853167852417281	598463.567322778\\
0.00148735210729351	585156.956774109\\
0.00259294379740467	567004.149732562\\
0.00452035365636025	547459.060457805\\
0.00788046281566991	530350.735229755\\
0.0137382379588326	517121.411340298\\
0.0239502661998749	506095.662655073\\
0.041753189365604	494553.388833757\\
0.0727895384398315	480395.521789577\\
0.126896100316792	462026.733881024\\
0.221221629107045	437748.125624411\\
0.385662042116347	405389.006484337\\
0.672335753649933	362127.503997546\\
1.17210229753348	304872.351390575\\
2.04335971785694	232722.085529138\\
3.56224789026244	152193.08450688\\
6.21016941891562	79412.8145559628\\
10.8263673387405	30610.8548569415\\
18.873918221351	8107.27674244073\\
32.9034456231267	1389.76194425982\\
57.3615251044868	147.563309829119\\
100	9.42128404589789\\
};
\addlegendentry{$\text{FS}_m$}

\addplot [color=mycolor2, line width=1.5pt]
  table[row sep=crcr]{%
0	596150.327237399\\
1e-05	596150.411807951\\
1.74332882219999e-05	596150.559185296\\
3.03919538231319e-05	596150.810304654\\
5.29831690628371e-05	596150.955191383\\
9.23670857187387e-05	596143.997541428\\
0.000161026202756094	596044.285973726\\
0.000280721620394118	595333.239196741\\
0.000489390091847749	592405.427142798\\
0.000853167852417281	584779.631050099\\
0.00148735210729351	571108.600279143\\
0.00259294379740467	553098.510946206\\
0.00452035365636025	535208.829877034\\
0.00788046281566991	521988.851643269\\
0.0137382379588326	514667.798421826\\
0.0239502661998749	510802.751394454\\
0.041753189365604	507211.486171539\\
0.0727895384398315	501674.053046591\\
0.126896100316792	492282.697260185\\
0.221221629107045	476443.163468774\\
0.385662042116347	450342.653903572\\
0.672335753649933	409016.49563372\\
1.17210229753348	347761.218972308\\
2.04335971785694	266061.174067648\\
3.56224789026244	173292.58935544\\
6.21016941891562	89637.2398372113\\
10.8263673387405	34121.1301159133\\
18.873918221351	8893.70251266404\\
32.9034456231267	1495.94593320649\\
57.3615251044868	155.482573868984\\
100	9.69948399414258\\
};
\addlegendentry{$\text{DC}_m$}

\addplot [color=mycolor3, dashed, line width=1.0pt, mark size=2.0pt, mark=o, mark options={solid, mycolor3}]
  table[row sep=crcr]{%
0	724215.786878658\\
1e-05	728919.126308805\\
1.74332882219999e-05	729367.140750971\\
3.03919538231319e-05	729670.446133379\\
5.29831690628371e-05	729864.480032926\\
9.23670857187387e-05	729851.970199102\\
0.000161026202756094	728852.289316295\\
0.000280721620394118	722060.061496346\\
0.000489390091847749	694859.638451205\\
0.000853167852417281	626049.711668105\\
0.00148735210729351	506333.119780999\\
0.00259294379740467	352793.588366412\\
0.00452035365636025	203392.41872035\\
0.00788046281566991	94739.1110861931\\
0.0137382379588326	37494.3681300986\\
0.0239502661998749	15545.3190972708\\
0.041753189365604	8516.711180901\\
0.0727895384398315	5816.48812334529\\
0.126896100316792	4284.92916571103\\
0.221221629107045	3226.11439526025\\
0.385662042116347	2450.4384726821\\
0.672335753649933	1867.83162296095\\
1.17210229753348	1411.50449169678\\
2.04335971785694	1018.53186532132\\
3.56224789026244	653.148183643274\\
6.21016941891562	340.07229195399\\
10.8263673387405	131.582045951573\\
18.873918221351	35.0329195984245\\
32.9034456231267	6.03735406170126\\
57.3615251044868	0.644190977318116\\
100	0.0413085569845386\\
};
\addlegendentry{$\text{FS}_f$}

\addplot [color=mycolor4, line width=1.0pt, mark size=2.0pt, mark=diamond, mark options={solid, mycolor4}]
  table[row sep=crcr]{%
0	707456.22977642\\
1e-05	707455.514070321\\
1.74332882219999e-05	707454.265901863\\
3.03919538231319e-05	707452.040157226\\
5.29831690628371e-05	707445.64517467\\
9.23670857187387e-05	707372.587256462\\
0.000161026202756094	706493.282241343\\
0.000280721620394118	700349.038066535\\
0.000489390091847749	675154.237576373\\
0.000853167852417281	609640.834621561\\
0.00148735210729351	492393.383444934\\
0.00259294379740467	338453.364294821\\
0.00452035365636025	186865.56743787\\
0.00788046281566991	77872.4609458807\\
0.0137382379588326	23739.841181635\\
0.0239502661998749	6204.80441230381\\
0.041753189365604	2656.3958707532\\
0.0727895384398315	2202.55876272054\\
0.126896100316792	2130.97446194787\\
0.221221629107045	2061.09632000345\\
0.385662042116347	1948.06477940086\\
0.672335753649933	1769.16129749562\\
1.17210229753348	1504.02016836352\\
2.04335971785694	1150.45902690369\\
3.56224789026244	749.120149787557\\
6.21016941891562	387.352216791925\\
10.8263673387405	147.386702815395\\
18.873918221351	38.3985191639651\\
32.9034456231267	6.45546591089634\\
57.3615251044868	0.670589934700131\\
100	0.0418095483858326\\
};
\addlegendentry{$\text{DC}_f$}

\end{axis}
\end{tikzpicture}%
\subcaption{Pressure}\label{fig:9a}
\end{minipage}%
\begin{minipage}[b]{0.5\textwidth}
\centering
\setlength\figureheight{5cm}
\setlength\figurewidth{5.5cm}
\definecolor{mycolor1}{rgb}{0.69804,0.09412,0.16863}%
\definecolor{mycolor2}{rgb}{0.12941,0.40000,0.67451}%
\begin{tikzpicture}

\begin{axis}[%
width=0.951\figurewidth,
height=\figureheight,
at={(0\figurewidth,0\figureheight)},
scale only axis,
xmode=log,
xmin=1e-05,
xmax=100,
xminorticks=true,
xlabel style={font=\fontsize{9}{144}\selectfont\color{white!5!black}},
xlabel={Time (s)},
ymin=-5.5e-05,
ymax=-4.0e-05,
ylabel style={font=\fontsize{9}{144}\selectfont\color{white!5!black}},
ylabel={Volumetric Strain},
ticklabel style={font=\fontsize{9}{144}},
axis background/.style={fill=white},
legend style={font=\fontsize{9}{144}\selectfont\color{white!5!black}, at={(1,1)}, anchor=north east, legend cell align=left, align=left, fill=none, draw=none}
]
\addplot [color=mycolor1, line width=1.5pt]
  table[row sep=crcr]{%
0	-4.18158640816238e-05\\
1e-05	-4.18161767907406e-05\\
1.74332882219999e-05	-4.18152745865571e-05\\
3.03919538231319e-05	-4.18139826315644e-05\\
5.29831690628371e-05	-4.18127482185501e-05\\
9.23670857187387e-05	-4.18119895839307e-05\\
0.000161026202756094	-4.18149973149767e-05\\
0.000280721620394118	-4.18428044551515e-05\\
0.000489390091847749	-4.1960122321609e-05\\
0.000853167852417281	-4.22627387492821e-05\\
0.00148735210729351	-4.27987450890163e-05\\
0.00259294379740467	-4.3504863338294e-05\\
0.00452035365636025	-4.42244975397733e-05\\
0.00788046281566991	-4.47980348981333e-05\\
0.0137382379588326	-4.51734290156381e-05\\
0.0239502661998749	-4.54209615609508e-05\\
0.041753189365604	-4.5639938067789e-05\\
0.0727895384398315	-4.58940706364892e-05\\
0.126896100316792	-4.62195795801702e-05\\
0.221221629107045	-4.66484613927117e-05\\
0.385662042116347	-4.7219923277403e-05\\
0.672335753649933	-4.79844894020049e-05\\
1.17210229753348	-4.89971911801544e-05\\
2.04335971785694	-5.02738827958219e-05\\
3.56224789026244	-5.16988938803775e-05\\
6.21016941891562	-5.29866321634253e-05\\
10.8263673387405	-5.38499888707825e-05\\
18.873918221351	-5.42480539252861e-05\\
32.9034456231267	-5.43668693537117e-05\\
57.3615251044868	-5.43888391905437e-05\\
100	-5.43912822855731e-05\\
};
\addlegendentry{FS}

\addplot [color=mycolor2, dashed, line width=1.5pt]
  table[row sep=crcr]{%
0	-4.26256686252483e-05\\
1e-05	-4.26256692283537e-05\\
1.74332882219999e-05	-4.26256702819908e-05\\
3.03919538231319e-05	-4.26256723546753e-05\\
5.29831690628371e-05	-4.26256878747709e-05\\
9.23670857187387e-05	-4.26260080477456e-05\\
0.000161026202756094	-4.26301264866006e-05\\
0.000280721620394118	-4.26591417653426e-05\\
0.000489390091847749	-4.27783216177017e-05\\
0.000853167852417281	-4.30884323139736e-05\\
0.00148735210729351	-4.36438160375036e-05\\
0.00259294379740467	-4.43740140643911e-05\\
0.00452035365636025	-4.50956221489712e-05\\
0.00788046281566991	-4.56203892423417e-05\\
0.0137382379588326	-4.58935476718378e-05\\
0.0239502661998749	-4.6006805789842e-05\\
0.041753189365604	-4.6076321725281e-05\\
0.0727895384398315	-4.61694467793441e-05\\
0.126896100316792	-4.63254280479889e-05\\
0.221221629107045	-4.65883647329958e-05\\
0.385662042116347	-4.70216279649787e-05\\
0.672335753649933	-4.77076337352365e-05\\
1.17210229753348	-4.87244586895305e-05\\
2.04335971785694	-5.00806623699771e-05\\
3.56224789026244	-5.16206011575429e-05\\
6.21016941891562	-5.30092617717762e-05\\
10.8263673387405	-5.39308168032862e-05\\
18.873918221351	-5.43495863142e-05\\
32.9034456231267	-5.44723873288983e-05\\
57.3615251044868	-5.44946386956976e-05\\
100	-5.44970586586441e-05\\
};
\addlegendentry{DC}

\end{axis}
\end{tikzpicture}%
\subcaption{Volumetric strain, $\text{E}^v$}\label{fig:9b}
\end{minipage}
\caption{Matrix and fracture continuum element averaged pressure (a) and total element volumetric strain (b) evolutions for the (deformable) anisotropic test case with one (vertical) fracture set. `$\text{FS}_\alpha$' and `$\text{DC}_\alpha$' denote quantities related to fine-scale and dual-continuum models for continuum $\alpha$ respectively.} \label{fig:9}
\end{figure}
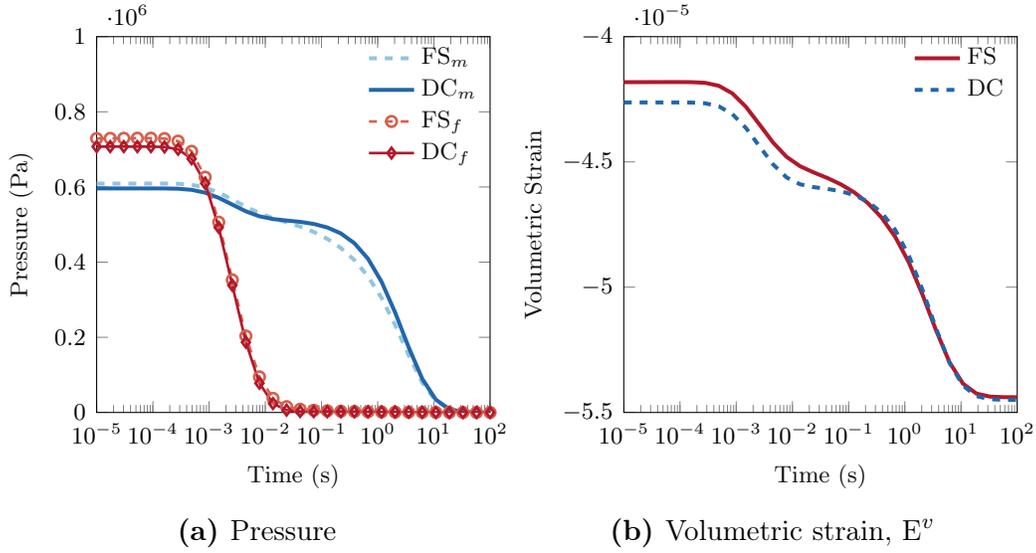

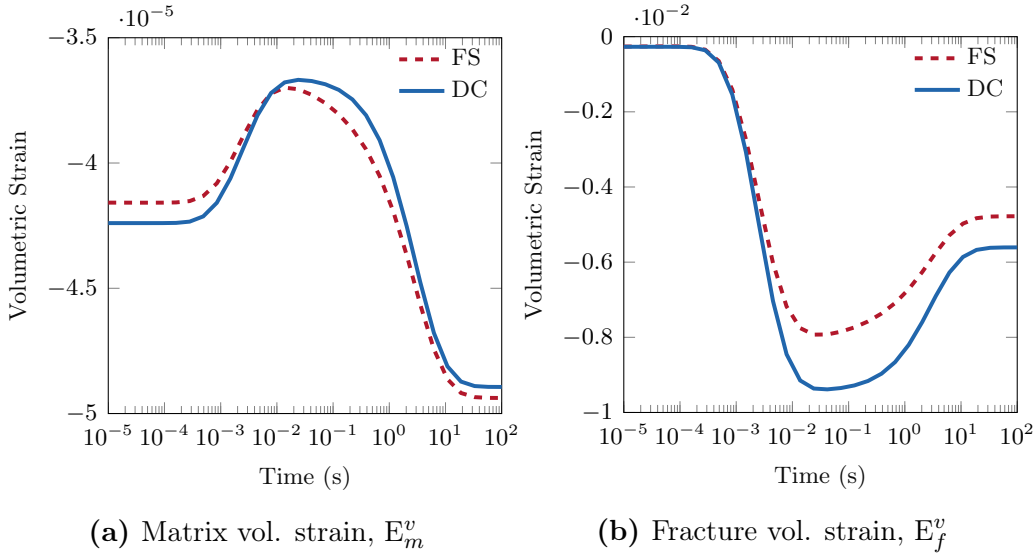
\begin{figure}[h]
\centering
\begin{minipage}[b]{0.5\textwidth}
\centering
\setlength\figureheight{5cm}
\setlength\figurewidth{5.5cm}
\definecolor{mycolor1}{rgb}{0.69804,0.09412,0.16863}%
\definecolor{mycolor2}{rgb}{0.12941,0.40000,0.67451}%
\begin{tikzpicture}

\begin{axis}[%
width=0.951\figurewidth,
height=\figureheight,
at={(0\figurewidth,0\figureheight)},
scale only axis,
xmode=log,
xmin=1e-05,
xmax=100,
xminorticks=true,
xlabel style={font=\fontsize{9}{144}\selectfont\color{white!5!black}},
xlabel={Time (s)},
ymin=-5e-05,
ymax=-3.5e-05,
ylabel style={font=\fontsize{9}{144}\selectfont\color{white!5!black}},
ylabel={Volumetric Strain},
ticklabel style={font=\fontsize{9}{144}},
axis background/.style={fill=white},
legend style={font=\fontsize{9}{144}\selectfont\color{white!5!black}, at={(1,1)}, anchor=north east, legend cell align=left, align=left, fill=none, draw=none}
]
\addplot [color=mycolor1, dashed, line width=1.5pt]
  table[row sep=crcr]{%
0	-4.1582448025675e-05\\
1e-05	-4.15813199186376e-05\\
1.74332882219999e-05	-4.15822593994552e-05\\
3.03919538231319e-05	-4.15834616912954e-05\\
5.29831690628371e-05	-4.15838908305969e-05\\
9.23670857187387e-05	-4.15823817257639e-05\\
0.000161026202756094	-4.15727558209598e-05\\
0.000280721620394118	-4.15193865599424e-05\\
0.000489390091847749	-4.13159767414046e-05\\
0.000853167852417281	-4.08119409796115e-05\\
0.00148735210729351	-3.99507557277956e-05\\
0.00259294379740467	-3.88750557238306e-05\\
0.00452035365636025	-3.78773665671235e-05\\
0.00788046281566991	-3.72267226469615e-05\\
0.0137382379588326	-3.69925678116372e-05\\
0.0239502661998749	-3.70554282268082e-05\\
0.041753189365604	-3.72797852046932e-05\\
0.0727895384398315	-3.76120820483147e-05\\
0.126896100316792	-3.80597913451386e-05\\
0.221221629107045	-3.86568270481743e-05\\
0.385662042116347	-3.94531298528034e-05\\
0.672335753649933	-4.05153813080975e-05\\
1.17210229753348	-4.19178906217158e-05\\
2.04335971785694	-4.36831102485965e-05\\
3.56224789026244	-4.56530777488726e-05\\
6.21016941891562	-4.74341036645324e-05\\
10.8263673387405	-4.86288341733657e-05\\
18.873918221351	-4.91799339972455e-05\\
32.9034456231267	-4.93444839798574e-05\\
57.3615251044868	-4.93749182077747e-05\\
100	-4.9378303205263e-05\\
};
\addlegendentry{FS}

\addplot [color=mycolor2, line width=1.5pt]
  table[row sep=crcr]{%
0	-4.2398683310047e-05\\
1e-05	-4.23986733309117e-05\\
1.74332882219999e-05	-4.23986559302984e-05\\
3.03919538231319e-05	-4.23986251833748e-05\\
5.29831690628371e-05	-4.23985507574128e-05\\
9.23670857187387e-05	-4.23979083506671e-05\\
0.000161026202756094	-4.23905619528941e-05\\
0.000280721620394118	-4.23395732814068e-05\\
0.000489390091847749	-4.21307841761325e-05\\
0.000853167852417281	-4.15881811587711e-05\\
0.00148735210729351	-4.06176639803006e-05\\
0.00259294379740467	-3.93448887956096e-05\\
0.00452035365636025	-3.80952972730659e-05\\
0.00788046281566991	-3.7205447210086e-05\\
0.0137382379588326	-3.67817188291325e-05\\
0.0239502661998749	-3.6680505642996e-05\\
0.041753189365604	-3.67278167873865e-05\\
0.0727895384398315	-3.68564179880686e-05\\
0.126896100316792	-3.70822656452085e-05\\
0.221221629107045	-3.74637461327643e-05\\
0.385662042116347	-3.80923761591978e-05\\
0.672335753649933	-3.90877162162637e-05\\
1.17210229753348	-4.05630492730302e-05\\
2.04335971785694	-4.25307953040007e-05\\
3.56224789026244	-4.47651285296017e-05\\
6.21016941891562	-4.67799709100333e-05\\
10.8263673387405	-4.81170798603014e-05\\
18.873918221351	-4.8724684642657e-05\\
32.9034456231267	-4.89028604393728e-05\\
57.3615251044868	-4.89351456847524e-05\\
100	-4.89386568946041e-05\\
};
\addlegendentry{DC}

\end{axis}
\end{tikzpicture}%
\subcaption{Matrix vol. strain, $\text{E}^v_m$}\label{fig:10a}
\end{minipage}%
\begin{minipage}[b]{0.5\textwidth}
\centering
\setlength\figureheight{5cm}
\setlength\figurewidth{5.5cm}
\definecolor{mycolor1}{rgb}{0.69804,0.09412,0.16863}%
\definecolor{mycolor2}{rgb}{0.12941,0.40000,0.67451}%
\begin{tikzpicture}

\begin{axis}[%
width=0.951\figurewidth,
height=\figureheight,
at={(0\figurewidth,0\figureheight)},
scale only axis,
xmode=log,
xmin=1e-05,
xmax=100,
xminorticks=true,
xlabel style={font=\fontsize{9}{144}\selectfont\color{white!5!black}},
xlabel={Time (s)},
ymin=-0.01,
ymax=0,
ylabel style={font=\fontsize{9}{144}\selectfont\color{white!5!black}},
ylabel={Volumetric Strain},
ticklabel style={font=\fontsize{9}{144}},
axis background/.style={fill=white},
legend style={font=\fontsize{9}{144}\selectfont\color{white!5!black}, at={(1,1)}, anchor=north east, legend cell align=left, align=left, fill=none, draw=none}
]
\addplot [color=mycolor1, dashed, line width=1.5pt]
  table[row sep=crcr]{%
0	-0.000261786274392389\\
1e-05	-0.000263144406808291\\
1.74332882219999e-05	-0.000261407907608764\\
3.03919538231319e-05	-0.000259056046850751\\
5.29831690628371e-05	-0.000257487087012938\\
9.23670857187387e-05	-0.000258193568675592\\
0.000161026202756094	-0.000270102467157596\\
0.000280721620394118	-0.000346630608455299\\
0.000489390091847749	-0.000649000486368224\\
0.000853167852417281	-0.00140948908217833\\
0.00148735210729351	-0.00272673317197387\\
0.00259294379740467	-0.00440661808823664\\
0.00452035365636025	-0.00602573677472754\\
0.00788046281566991	-0.00717997412941302\\
0.0137382379588326	-0.00775478615649218\\
0.0239502661998749	-0.00792906800760696\\
0.041753189365604	-0.007924216448503\\
0.0727895384398315	-0.00785080886334539\\
0.126896100316792	-0.00773597322062061\\
0.221221629107045	-0.00757793451057439\\
0.385662042116347	-0.00736661673796594\\
0.672335753649933	-0.00708684377178702\\
1.17210229753348	-0.00672050332311125\\
2.04335971785694	-0.00626139306046152\\
3.56224789026244	-0.00574925320181039\\
6.21016941891562	-0.00528566853658255\\
10.8263673387405	-0.00497424647324597\\
18.873918221351	-0.00483042514913008\\
32.9034456231267	-0.00478744389327653\\
57.3615251044868	-0.00477948905289576\\
100	-0.00477860385066779\\
};
\addlegendentry{FS}

\addplot [color=mycolor2, line width=1.5pt]
  table[row sep=crcr]{%
0	-0.000269383998511367\\
1e-05	-0.000269394570772851\\
1.74332882219999e-05	-0.00026941300762269\\
3.03919538231319e-05	-0.000269445796483841\\
5.29831690628371e-05	-0.000269535668115462\\
9.23670857187387e-05	-0.000270497605429222\\
0.000161026202756094	-0.000281955095659434\\
0.000280721620394118	-0.000361908057217241\\
0.000489390091847749	-0.000689668225745373\\
0.000853167852417281	-0.00154183933636124\\
0.00148735210729351	-0.00306676972118325\\
0.00259294379740467	-0.00506847015757708\\
0.00452035365636025	-0.00703842017317835\\
0.00788046281566991	-0.00845214747946581\\
0.0137382379588326	-0.00914861056153446\\
0.0239502661998749	-0.00936298065248898\\
0.041753189365604	-0.00938523275468186\\
0.0727895384398315	-0.00934988520926362\\
0.126896100316792	-0.00928024466842555\\
0.221221629107045	-0.00916208234636431\\
0.385662042116347	-0.00896734418194008\\
0.672335753649933	-0.00865900523518908\\
1.17210229753348	-0.00820197246577326\\
2.04335971785694	-0.00759239786128037\\
3.56224789026244	-0.00690023775647085\\
6.21016941891562	-0.00627607083265291\\
10.8263673387405	-0.00586185402284503\\
18.873918221351	-0.00567362635618569\\
32.9034456231267	-0.00561842974996482\\
57.3615251044868	-0.00560842815662999\\
100	-0.00560734042093466\\
};
\addlegendentry{DC}

\end{axis}
\end{tikzpicture}%
\subcaption{Fracture vol. strain, $\text{E}^v_f$}\label{fig:10b}
\end{minipage}
\caption{Matrix (a) and fracture (b) volumetric strain evolutions for the (deformable) anisotropic test case with one (vertical) fracture set. `$\text{FS}$' and `$\text{DC}$' denote quantities related to fine-scale and dual-continuum models.} \label{fig:10}
\end{figure}

Here we show the results for the geometry-induced (single-fracture set) anisotropy case. Pressure and total volumetric strain are given in \cref{fig:9}, whilst \cref{fig:10} shows the individual continuum volumetric strains.

\cref{fig:9a} now shows a smaller disparity between the initial matrix and fracture pressures, with the fracture pressure being only slightly higher. Further, we do not observe the Mandel-Cryer effect in the FS model. However, away from the initial pressures, the general trends we see in \cref{fig:6a} can still be observed in \cref{fig:9a}. Specifically, a rapid decrease in fracture pressure followed by a smoother matrix pressure decrease. As expected, the late-time differences in matrix pressure observed previously are present in the current test. For both modelling approaches there is a good agreement in matrix and fracture pressure evolutions. The total element volumetric strain evolutions are also similar between the two modelling approaches, with an overall increase in strain as the material compacts. 

The similarity in total volumetric strain between the two approaches is reflected in the individual continuum strains (\cref{fig:10}). The matrix shows early-time expansion behaviour followed by contraction. Fracture deformation is coupled to matrix deformation (and vice versa). Fracture contraction is therefore followed by a period of expansion as the matrix drains and contracts.  

The small difference in initial continuum pressures observed in \cref{fig:9a} can be explained by considering the geometric anisotropy induced by the fractures. With the fracture set being aligned with the direction of loading (\cref{fig:3}a), the stiffer matrix acts like a series of columns, supporting a significant portion of the applied load. Through the coupling between stress and pressure, the low portion of stress `seen' by the fracture phase leads to the low induced fracture pressure shown in \cref{fig:9a}. Finally, the absence of the Mandel-Cryer effect in the current case is due to the pressure being at equilibrium within the single fracture set. As a result local processes do not drive early-time poroelastic intra- and inter-continuum pressure generation. 

\subsubsection{Material-induced anisotropy: numerical computation of $\mathbb{C}^*$}
\begin{figure}[h]
\centering
\begin{minipage}[b]{0.5\textwidth}
\centering
\setlength\figureheight{5cm}
\setlength\figurewidth{5.5cm}
\definecolor{mycolor1}{rgb}{0.57255,0.77255,0.87059}%
\definecolor{mycolor2}{rgb}{0.12941,0.40000,0.67451}%
\definecolor{mycolor3}{rgb}{0.83922,0.37647,0.30196}%
\definecolor{mycolor4}{rgb}{0.69804,0.09412,0.16863}%
\begin{tikzpicture}

\begin{axis}[%
width=0.951\figurewidth,
height=\figureheight,
at={(0\figurewidth,0\figureheight)},
scale only axis,
xmode=log,
xmin=1e-05,
xmax=100,
xminorticks=true,
xlabel style={font=\fontsize{9}{144}\selectfont\color{white!5!black}},
xlabel={Time (s)},
ymin=0,
ymax=2500000,
ylabel style={font=\fontsize{9}{144}\selectfont\color{white!5!black}},
ylabel={Pressure (Pa)},
ticklabel style={font=\fontsize{9}{144}},
axis background/.style={fill=white},
legend style={font=\fontsize{9}{144}\selectfont\color{white!5!black}, at={(1,1)}, anchor=north east, legend cell align=left, align=left, fill=none, draw=none}
]
\addplot [color=mycolor1, dashed, line width=1.5pt]
  table[row sep=crcr]{%
0	609914.761570563\\
1e-05	613227.92006033\\
1.74332882219999e-05	616024.496567269\\
3.03919538231319e-05	619248.423381277\\
5.29831690628371e-05	623294.387115031\\
9.23670857187387e-05	628399.999996861\\
0.000161026202756094	634472.052885645\\
0.000280721620394118	641160.860163987\\
0.000489390091847749	648513.075286269\\
0.000853167852417281	657436.40301007\\
0.00148735210729351	669145.102782996\\
0.00259294379740467	684367.378745029\\
0.00452035365636025	703463.904656611\\
0.00788046281566991	727159.772910653\\
0.0137382379588326	756865.292818378\\
0.0239502661998749	794518.672232293\\
0.041753189365604	842399.338347721\\
0.0727895384398315	902917.029305904\\
0.126896100316792	978422.678373768\\
0.221221629107045	1070723.92552373\\
0.385662042116347	1179392.03566398\\
0.672335753649933	1295704.33071889\\
1.17210229753348	1388351.47563366\\
2.04335971785694	1392626.05892089\\
3.56224789026244	1237488.20826755\\
6.21016941891562	917594.537640808\\
10.8263673387405	534559.97656218\\
18.873918221351	229923.443788031\\
32.9034456231267	68540.2460203535\\
57.3615251044868	13366.0034024952\\
100	1628.55496794574\\
};
\addlegendentry{$\text{FS}_m$}

\addplot [color=mycolor2, line width=1.5pt]
  table[row sep=crcr]{%
0	625694.740088818\\
1e-05	625699.050176361\\
1.74332882219999e-05	625706.564025392\\
3.03919538231319e-05	625719.662980552\\
5.29831690628371e-05	625742.498297317\\
9.23670857187387e-05	625782.306336936\\
0.000161026202756094	625851.700505344\\
0.000280721620394118	625972.664187524\\
0.000489390091847749	626183.503637334\\
0.000853167852417281	626550.944527207\\
0.00148735210729351	627191.145468353\\
0.00259294379740467	628306.104813854\\
0.00452035365636025	630246.443364795\\
0.00788046281566991	633618.783002059\\
0.0137382379588326	639466.716420818\\
0.0239502661998749	649567.746991871\\
0.041753189365604	666896.509586318\\
0.0727895384398315	696276.662103637\\
0.126896100316792	745091.594996916\\
0.221221629107045	823428.506746371\\
0.385662042116347	941663.264194743\\
0.672335753649933	1100458.37625783\\
1.17210229753348	1267436.7715911\\
2.04335971785694	1356050.84419178\\
3.56224789026244	1260924.34089251\\
6.21016941891562	960368.395256154\\
10.8263673387405	568552.007028742\\
18.873918221351	247675.367418079\\
32.9034456231267	74802.3391404834\\
57.3615251044868	14797.8347612899\\
100	1831.17867562291\\
};
\addlegendentry{$\text{DC}_m$}

\addplot [color=mycolor3, dashed, line width=1.0pt, mark size=2.0pt, mark=o, mark options={solid, mycolor3}]
  table[row sep=crcr]{%
0	1343929.73157307\\
1e-05	1414035.83028623\\
1.74332882219999e-05	1471580.40127443\\
3.03919538231319e-05	1536714.47596811\\
5.29831690628371e-05	1616554.62490736\\
9.23670857187387e-05	1711621.90569891\\
0.000161026202756094	1810667.84538459\\
0.000280721620394118	1892339.91622415\\
0.000489390091847749	1941179.86818147\\
0.000853167852417281	1960834.32576978\\
0.00148735210729351	1965561.02848595\\
0.00259294379740467	1965280.6281017\\
0.00452035365636025	1963031.81885029\\
0.00788046281566991	1959063.07797338\\
0.0137382379588326	1953176.94579213\\
0.0239502661998749	1945093.23256094\\
0.041753189365604	1934409.43196963\\
0.0727895384398315	1920645.84165799\\
0.126896100316792	1903287.56814509\\
0.221221629107045	1881561.16796244\\
0.385662042116347	1852104.94069713\\
0.672335753649933	1800620.89082914\\
1.17210229753348	1693224.39687186\\
2.04335971785694	1490888.70897927\\
3.56224789026244	1184807.43014158\\
6.21016941891562	813626.354616059\\
10.8263673387405	454894.380778905\\
18.873918221351	192295.456157856\\
32.9034456231267	56981.9472028295\\
57.3615251044868	11092.3677847779\\
100	1350.88697606738\\
};
\addlegendentry{$\text{FS}_f$}

\addplot [color=mycolor4, line width=1.0pt, mark size=2.0pt, mark=diamond, mark options={solid, mycolor4}]
  table[row sep=crcr]{%
0	1984444.33036156\\
1e-05	1984443.3694697\\
1.74332882219999e-05	1984441.69433055\\
3.03919538231319e-05	1984438.7740466\\
5.29831690628371e-05	1984433.68313599\\
9.23670857187387e-05	1984424.80832262\\
0.000161026202756094	1984409.33757054\\
0.000280721620394118	1984382.36989767\\
0.000489390091847749	1984335.36529502\\
0.000853167852417281	1984253.44791322\\
0.00148735210729351	1984110.72128548\\
0.00259294379740467	1983862.15173147\\
0.00452035365636025	1983429.57139935\\
0.00788046281566991	1982677.73874487\\
0.0137382379588326	1981373.99158628\\
0.0239502661998749	1979122.04267254\\
0.041753189365604	1975258.68444384\\
0.0727895384398315	1968707.72893402\\
0.126896100316792	1957799.05462922\\
0.221221629107045	1939867.29296274\\
0.385662042116347	1908877.58984891\\
0.672335753649933	1847507.34357213\\
1.17210229753348	1721587.24512675\\
2.04335971785694	1499660.83603758\\
3.56224789026244	1183654.54539094\\
6.21016941891562	813131.877165222\\
10.8263673387405	457622.349551757\\
18.873918221351	195461.36051136\\
32.9034456231267	58659.1139821215\\
57.3615251044868	11583.644790946\\
100	1432.78502382925\\
};
\addlegendentry{$\text{DC}_f$}

\end{axis}
\end{tikzpicture}%
\subcaption{Pressure}\label{fig:11a}
\end{minipage}%
\begin{minipage}[b]{0.5\textwidth}
\centering
\setlength\figureheight{5cm}
\setlength\figurewidth{5.5cm}
\definecolor{mycolor1}{rgb}{0.69804,0.09412,0.16863}%
\definecolor{mycolor2}{rgb}{0.12941,0.40000,0.67451}%
\begin{tikzpicture}

\begin{axis}[%
width=0.951\figurewidth,
height=\figureheight,
at={(0\figurewidth,0\figureheight)},
scale only axis,
xmode=log,
xmin=1e-05,
xmax=100,
xminorticks=true,
xlabel style={font=\fontsize{9}{144}\selectfont\color{white!5!black}},
xlabel={Time (s)},
ymin=-6e-04,
ymax=0,
ylabel style={font=\fontsize{9}{144}\selectfont\color{white!5!black}},
ylabel={Volumetric Strain},
ticklabel style={font=\fontsize{9}{144}},
axis background/.style={fill=white},
legend style={font=\fontsize{9}{144}\selectfont\color{white!5!black}, at={(1,1)}, anchor=north east, legend cell align=left, align=left, fill=none, draw=none}
]
\addplot [color=mycolor1, dashed, line width=1.5pt]
  table[row sep=crcr]{%
0	-4.25008067768869e-05\\
1e-05	-4.27596106844917e-05\\
1.74332882219999e-05	-4.29752655130472e-05\\
3.03919538231319e-05	-4.32206588151728e-05\\
5.29831690628371e-05	-4.35234374059078e-05\\
9.23670857187387e-05	-4.38948118625334e-05\\
0.000161026202756094	-4.43136645054018e-05\\
0.000280721620394118	-4.47314523623269e-05\\
0.000489390091847749	-4.51254954501848e-05\\
0.000853167852417281	-4.55404396461802e-05\\
0.00148735210729351	-4.60512802115725e-05\\
0.00259294379740467	-4.67054474489178e-05\\
0.00452035365636025	-4.75237466261466e-05\\
0.00788046281566991	-4.85384029404889e-05\\
0.0137382379588326	-4.98106984213173e-05\\
0.0239502661998749	-5.14247664504052e-05\\
0.041753189365604	-5.34791519711045e-05\\
0.0727895384398315	-5.6077153315272e-05\\
0.126896100316792	-5.93232836536969e-05\\
0.221221629107045	-6.33904968326194e-05\\
0.385662042116347	-6.91538427677459e-05\\
0.672335753649933	-8.05181333200747e-05\\
1.17210229753348	-0.000107049631073884\\
2.04335971785694	-0.00016038171958048\\
3.56224789026244	-0.000244232169969347\\
6.21016941891562	-0.000348443466452701\\
10.8263673387405	-0.000450550643879066\\
18.873918221351	-0.000525747497019725\\
32.9034456231267	-0.000564577308118061\\
57.3615251044868	-0.000577754101927596\\
100	-0.000580551748191156\\
};
\addlegendentry{FS}

\addplot [color=mycolor2, line width=1.5pt]
  table[row sep=crcr]{%
0	-4.56963416950848e-05\\
1e-05	-4.56965213751231e-05\\
1.74332882219999e-05	-4.5696834614396e-05\\
3.03919538231319e-05	-4.56973806870155e-05\\
5.29831690628371e-05	-4.56983326516006e-05\\
9.23670857187387e-05	-4.5699992179481e-05\\
0.000161026202756094	-4.57028851016932e-05\\
0.000280721620394118	-4.57079278676292e-05\\
0.000489390091847749	-4.57167173994263e-05\\
0.000853167852417281	-4.57320353776542e-05\\
0.00148735210729351	-4.57587242632424e-05\\
0.00259294379740467	-4.5805205048485e-05\\
0.00452035365636025	-4.58860946034984e-05\\
0.00788046281566991	-4.60266822393765e-05\\
0.0137382379588326	-4.6270474425624e-05\\
0.0239502661998749	-4.66915749241175e-05\\
0.041753189365604	-4.74140022601628e-05\\
0.0727895384398315	-4.86390604338585e-05\\
0.126896100316792	-5.06809862881848e-05\\
0.221221629107045	-5.40716886563443e-05\\
0.385662042116347	-6.02389970981564e-05\\
0.672335753649933	-7.380359263656e-05\\
1.17210229753348	-0.000104483927546327\\
2.04335971785694	-0.00016224436557988\\
3.56224789026244	-0.000248459369394658\\
6.21016941891562	-0.000352937453986244\\
10.8263673387405	-0.000455037108767012\\
18.873918221351	-0.0005309005651852\\
32.9034456231267	-0.000570585437962422\\
57.3615251044868	-0.000584250793397895\\
100	-0.000587197947704988\\
};
\addlegendentry{DC}

\end{axis}
\end{tikzpicture}%
\subcaption{Volumetric strain, $\text{E}^v$}\label{fig:11b}
\end{minipage}
\caption{Matrix and fracture continuum element averaged pressure (a) and total element volumetric strain (b) evolutions for the (deformable) anisotropic test case with two orthogonal fracture sets. `$\text{FS}_\alpha$' and `$\text{DC}_\alpha$' denote quantities related to fine-scale and dual-continuum models for continuum $\alpha$ respectively.} \label{fig:11}
\end{figure}
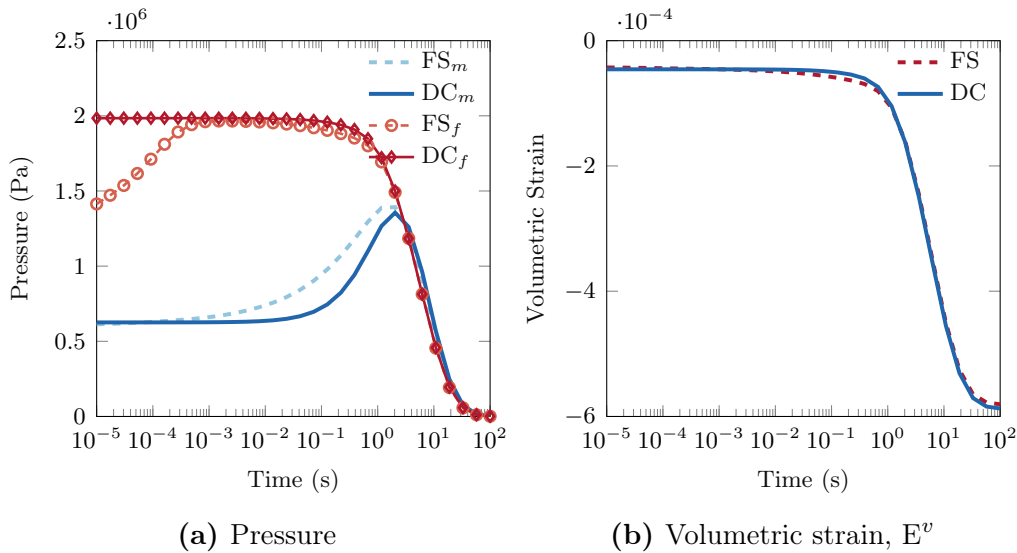

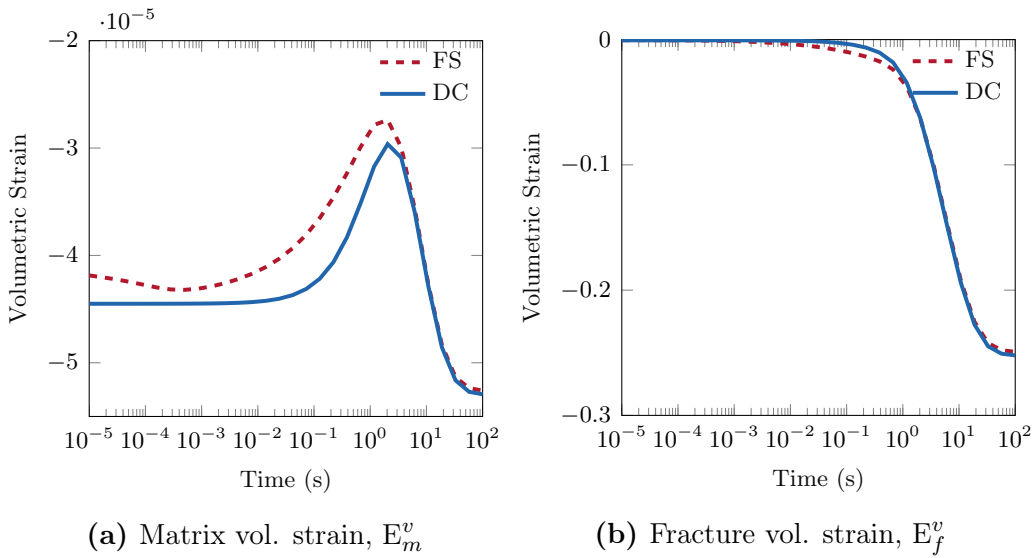
\begin{figure}[h]
\centering
\begin{minipage}[b]{0.5\textwidth}
\centering
\setlength\figureheight{5cm}
\setlength\figurewidth{5.5cm}
\definecolor{mycolor1}{rgb}{0.69804,0.09412,0.16863}%
\definecolor{mycolor2}{rgb}{0.12941,0.40000,0.67451}%
\begin{tikzpicture}

\begin{axis}[%
width=0.951\figurewidth,
height=\figureheight,
at={(0\figurewidth,0\figureheight)},
scale only axis,
xmode=log,
xmin=1e-05,
xmax=100,
xminorticks=true,
xlabel style={font=\fontsize{9}{144}\selectfont\color{white!5!black}},
xlabel={Time (s)},
ymin=-5.5e-05,
ymax=-2.0e-05,
ylabel style={font=\fontsize{9}{144}\selectfont\color{white!5!black}},
ylabel={Volumetric Strain},
ticklabel style={font=\fontsize{9}{144}},
axis background/.style={fill=white},
legend style={font=\fontsize{9}{144}\selectfont\color{white!5!black}, at={(1,1)}, anchor=north east, legend cell align=left, align=left, fill=none, draw=none}
]
\addplot [color=mycolor1, dashed, line width=1.5pt]
  table[row sep=crcr]{%
0	-4.1644299061668e-05\\
1e-05	-4.18634405327977e-05\\
1.74332882219999e-05	-4.20412795276731e-05\\
3.03919538231319e-05	-4.22370502929936e-05\\
5.29831690628371e-05	-4.24677828246405e-05\\
9.23670857187387e-05	-4.27312403600468e-05\\
0.000161026202756094	-4.29890379823991e-05\\
0.000280721620394118	-4.31683858253308e-05\\
0.000489390091847749	-4.32093052548146e-05\\
0.000853167852417281	-4.3107025352029e-05\\
0.00148735210729351	-4.28916989068856e-05\\
0.00259294379740467	-4.25870073693762e-05\\
0.00452035365636025	-4.21976631657262e-05\\
0.00788046281566991	-4.17104009700276e-05\\
0.0137382379588326	-4.10966322031383e-05\\
0.0239502661998749	-4.03170442244778e-05\\
0.041753189365604	-3.93252307361262e-05\\
0.0727895384398315	-3.80721371058595e-05\\
0.126896100316792	-3.65101037061882e-05\\
0.221221629107045	-3.46023780522702e-05\\
0.385662042116347	-3.23528344062431e-05\\
0.672335753649933	-2.99118253994394e-05\\
1.17210229753348	-2.78468682978852e-05\\
2.04335971785694	-2.73859727373219e-05\\
3.56224789026244	-2.99385232214202e-05\\
6.21016941891562	-3.56810385252248e-05\\
10.8263673387405	-4.27143435076061e-05\\
18.873918221351	-4.83496959279492e-05\\
32.9034456231267	-5.13420170161232e-05\\
57.3615251044868	-5.23657228888903e-05\\
100	-5.25835382471885e-05\\
};
\addlegendentry{FS}

\addplot [color=mycolor2, line width=1.5pt]
  table[row sep=crcr]{%
0	-4.45039638485809e-05\\
1e-05	-4.45038794714022e-05\\
1.74332882219999e-05	-4.45037323752294e-05\\
3.03919538231319e-05	-4.45034759412546e-05\\
5.29831690628371e-05	-4.45030289017027e-05\\
9.23670857187387e-05	-4.45022495926884e-05\\
0.000161026202756094	-4.45008910856751e-05\\
0.000280721620394118	-4.44985230190992e-05\\
0.000489390091847749	-4.44943954839095e-05\\
0.000853167852417281	-4.44872022133463e-05\\
0.00148735210729351	-4.44746692080747e-05\\
0.00259294379740467	-4.44528420119706e-05\\
0.00452035365636025	-4.44148566352514e-05\\
0.00788046281566991	-4.43488374371075e-05\\
0.0137382379588326	-4.42343543375103e-05\\
0.0239502661998749	-4.4036609714161e-05\\
0.041753189365604	-4.36973700353988e-05\\
0.0727895384398315	-4.31222026954287e-05\\
0.126896100316792	-4.21665225502682e-05\\
0.221221629107045	-4.06321130508376e-05\\
0.385662042116347	-3.8309212748649e-05\\
0.672335753649933	-3.51542255044175e-05\\
1.17210229753348	-3.17271842480179e-05\\
2.04335971785694	-2.96319029550203e-05\\
3.56224789026244	-3.08928412075812e-05\\
6.21016941891562	-3.59965224391415e-05\\
10.8263673387405	-4.28772706792243e-05\\
18.873918221351	-4.85639164330898e-05\\
32.9034456231267	-5.16355230687925e-05\\
57.3615251044868	-5.27024116324385e-05\\
100	-5.29329996617936e-05\\
};
\addlegendentry{DC}

\end{axis}
\end{tikzpicture}%
\subcaption{Matrix vol. strain, $\text{E}^v_m$}\label{fig:12a}
\end{minipage}%
\begin{minipage}[b]{0.5\textwidth}
\centering
\setlength\figureheight{5cm}
\setlength\figurewidth{5.5cm}
\definecolor{mycolor1}{rgb}{0.69804,0.09412,0.16863}%
\definecolor{mycolor2}{rgb}{0.12941,0.40000,0.67451}%
\begin{tikzpicture}

\begin{axis}[%
width=0.951\figurewidth,
height=\figureheight,
at={(0\figurewidth,0\figureheight)},
scale only axis,
xmode=log,
xmin=1e-05,
xmax=100,
xminorticks=true,
xlabel style={font=\fontsize{9}{144}\selectfont\color{white!5!black}},
xlabel={Time (s)},
ymin=-0.3,
ymax=0,
ylabel style={font=\fontsize{9}{144}\selectfont\color{white!5!black}},
ylabel={Volumetric Strain},
ticklabel style={font=\fontsize{9}{144}},
axis background/.style={fill=white},
legend style={font=\fontsize{9}{144}\selectfont\color{white!5!black}, at={(1,1)}, anchor=north east, legend cell align=left, align=left, fill=none, draw=none}
]
\addplot [color=mycolor1, dashed, line width=1.5pt]
  table[row sep=crcr]{%
0	-0.00044587161275441\\
1e-05	-0.000464809371543264\\
1.74332882219999e-05	-0.000482834326906048\\
3.03919538231319e-05	-0.000506449366867611\\
5.29831690628371e-05	-0.000540682110601096\\
9.23670857187387e-05	-0.000591876769795754\\
0.000161026202756094	-0.00066814420182435\\
0.000280721620394118	-0.000780854895994692\\
0.000489390091847749	-0.00094755190653653\\
0.000853167852417281	-0.00119155263272745\\
0.00148735210729351	-0.00153405055304633\\
0.00259294379740467	-0.00198627757825018\\
0.00452035365636025	-0.00255583340917751\\
0.00788046281566991	-0.00326417395309684\\
0.0137382379588326	-0.00415368489746459\\
0.0239502661998749	-0.00528258706333841\\
0.041753189365604	-0.00671924352846174\\
0.0727895384398315	-0.00853550795233955\\
0.126896100316792	-0.010803150360547\\
0.221221629107045	-0.013621104535682\\
0.385662042116347	-0.0174005241734729\\
0.672335753649933	-0.0239134700738361\\
1.17210229753348	-0.0374074518211883\\
2.04335971785694	-0.0627944953912785\\
3.56224789026244	-0.10116544918724\\
6.21016941891562	-0.147643360244114\\
10.8263673387405	-0.19252034086834\\
18.873918221351	-0.225355404459387\\
32.9034456231267	-0.242271837323851\\
57.3615251044868	-0.248008489562265\\
100	-0.249226254047765\\
};
\addlegendentry{FS}

\addplot [color=mycolor2, line width=1.5pt]
  table[row sep=crcr]{%
0	-0.000606946344274909\\
1e-05	-0.0006070708151888\\
1.74332882219999e-05	-0.000607287807453831\\
3.03919538231319e-05	-0.000607666091866193\\
5.29831690628371e-05	-0.000608325552438409\\
9.23670857187387e-05	-0.000609475167891055\\
0.000161026202756094	-0.000611479200528167\\
0.000280721620394118	-0.000614972508174731\\
0.000489390091847749	-0.000621061336765415\\
0.000853167852417281	-0.000631672657075551\\
0.00148735210729351	-0.000650161015985282\\
0.00259294379740467	-0.000682359934707417\\
0.00452035365636025	-0.000738395030336645\\
0.00788046281566991	-0.000835785064922492\\
0.0137382379588326	-0.00100466835816474\\
0.0239502661998749	-0.00129637868988234\\
0.041753189365604	-0.00179682577794294\\
0.0727895384398315	-0.00264541358874721\\
0.126896100316792	-0.00405842300269957\\
0.221221629107045	-0.00638005456847847\\
0.385662042116347	-0.0103825471134596\\
0.672335753649933	-0.0182659877784017\\
1.17210229753348	-0.0343509457211863\\
2.04335971785694	-0.0625826803109075\\
3.56224789026244	-0.102656613684168\\
6.21016941891562	-0.149536435931449\\
10.8263673387405	-0.194457895236616\\
18.873918221351	-0.227565851063654\\
32.9034456231267	-0.24483933122761\\
57.3615251044868	-0.250783071168922\\
100	-0.252064700944564\\
};
\addlegendentry{DC}

\end{axis}
\end{tikzpicture}%
\subcaption{Fracture vol. strain, $\text{E}^v_f$}\label{fig:12b}
\end{minipage}
\caption{Matrix (a) and fracture (b) volumetric strain evolutions for the (deformable) anisotropic test case with two orthogonal fracture sets. `$\text{FS}$' and `$\text{DC}$' denote quantities related to fine-scale and dual-continuum models.} \label{fig:12}
\end{figure}

\cref{fig:11} and \cref{fig:12} show pressure and total strain, and individual continuum strain results respectively, for the material-induced anisotropy case. Both models in \cref{fig:11a} show a strong difference in the early-time magnitudes of matrix and fracture pressures. The FS model shows similar early-time Mandel-Cryer fracture behaviour to what we observed in \cref{fig:6a}. In contrast, the FS matrix non-monotinicity is negligible in \cref{fig:11a} compared to the isotropic case. At later times we see a significant non-montonic evolution in matrix pressure that is shown by \textit{both} modelling approaches. This non-monotonic matrix pressure rise starts earlier in the FS model than the DC model. Finally, we observe that matrix and fracture diffusion starts at similar times, indicating a single-continuum response. Coupled to the delayed fracture diffusion response is the delayed increase in total volumetric strain (\cref{fig:11b}).  

In \cref{fig:12} both modelling approaches give similar continuum strain evolutions. Similar to \cref{fig:7a}, \cref{fig:12a} shows the DC approach misses the early-time matrix strain non-monotinicity displayed by the FS approach. However, contrast to \cref{fig:7a}, \cref{fig:12a} shows a smoother early-time FS matrix strain non-monotinicity, whilst the late-time matrix strain non-monotinicity for both approaches is much sharper.   

Results in \crefrange{fig:11}{fig:12} can be explained by considering the material anisotropy in the fracture continuum. The smoother early-time non-monotinicity in FS matrix strain occurs because the vertical fractures are stiffer. These fractures then expand less with incoming fluid, reducing poroelastic coupling (and thus deformation) with the matrix compared to the isotropic case. As a result, since FS matrix pressure does not change significantly, the initial matrix pressures for the two modelling approaches are similar. This result suggests mechanical anisotropy can noticeably affect the degree of inter-continuum coupling. The delay in fracture pressure diffusion occurs due to the low vertical fracture permeability. Accordingly, we see the non-monotonic rise in matrix pressure with local inter-continuum equilibration processes occuring at similar timescales to macroscopic fracture flow. Further, the pseudosteady state mass transfer assumption leads to the delayed response of this non-monotinicity in the DC model. The influx of fluid from the fractures into the matrix is accompanied by expansion of the matrix material, followed by contraction as fluid drains out (\cref{fig:12a}). 

The results in the current test show again how the DC model misses early-time effects due to local equilibration processes. Neglecting these local processes is implicit due to the steady state pressure assumption made during homogenisation. However, once local equilibration is reached, the DC model does predict the general poroleastic behaviours of the FS model. 

\section{Conclusions}
Dual-continuum models are an implicit approach to modelling multiscale materials. Further, with the appropriate extensions, they can be used to model complex multiphysics problems such as the coupled mechanics and flow phenomena studied in this work. 

In this paper we derived a dual-continuum poroelastic constitutive model that makes no assumptions on the material symmetry and mechanical properties of the dual-material and its constituents. We termed the resulting model the anisotropic, dual-stiffness constitutive model. Further, we discussed how under isotropy of the continua and bulk material, and void space assumptions of the high permeability transport phase, previously introduced constitutive models can be recovered from the constitutive model developed herein. 

Secondly, using numerical modelling, we investigated whether the dual-continuum approach with the derived constitutive model, is able to capture the global poroelastic behaviours of fine-scale explicit models. We introduced the computational framework used to carry out our investigation and described the resulting numerical tests therein. We observed that anisotropy can have measurable impacts on flow and deformation behaviour. However, we showed that the DC approach is capable of capturing the global poroelastic behaviours for both isotropic and anisotropic FS equivalents. Discrepancies between the two model representations arise when local equilibration processes not accounted for in the homogenisation approach, are significant.  

Finally, interesting extensions to the current work involve the study of non-linear poromechanical effects, and measurement methodologies for the material parameters used herein. In the former, it is well known that fracture (and soil aggregate) deformation is geometrically non-linear leading to coupled material non-linearities at the macroscopic scale. Modelling these scale effects requires comprehensive multiscale modelling approaches, and is an active area of research (\citealt{borja2016}; \citealt{Wang2018}; \citealt{Castelletto2019}). Lastly, in analogy to the work of \cite{Biot1957}, it highly desirable to develop methods of measurement for the parameters introduced in this work. In particular, the challenge remains how to map individual fracture characteristics to continuum properties. In this context, a microporomechanics framework could provide useful insights into experimental, and theoretical methodologies (e.g. \citealt{Lemarchand2009}).

\subsection*{Acknowledgements}
The authors are very grateful for the funding provided to them by the Natural Environmental Research Council to carry out this work, and to the reviewers who have helped to improve it.

\newpage
\bibliographystyle{agsm}

\end{document}